\documentclass[preprint, 12pt]{elsarticle}

\usepackage{amsmath, amssymb, amsthm}
\usepackage{color}
\usepackage{latexsym}
\usepackage{indentfirst}
\usepackage{graphicx}
\usepackage{placeins}
\usepackage{booktabs}
\usepackage{algorithm}
\usepackage{algorithmic}
\usepackage{multirow}
\usepackage{subfig}
\usepackage{fullpage}

\graphicspath{ {figure/} }

\theoremstyle{plain}

\theoremstyle{definition}

\theoremstyle{remark}

\DeclareMathOperator{\Tr}{Tr}

\DeclareMathOperator{\diag}{diag}
\DeclareMathOperator{\spanop}{span}

\newcommand{\etal}{\textit{et al}{}}
\newcommand{\ie}{\textit{i.e.}{~}}

\newcommand{\ud}{\,\mathrm{d}}

\newcommand{\TT}{\mathrm{T}}

\newcommand{\wt}[1]{\widetilde{#1}}

\newcommand{\mc}[1]{\mathcal{#1}}

\newcommand{\abs}[1]{\left\lvert#1\right\rvert}

\newcommand{\average}[1]{\left\langle#1\right\rangle}
\providecommand{\averageop}[3]{\left\langle#1\left\lvert#2\right\rvert#3\right\rangle}
\newcommand{\bra}[1]{\langle#1\rvert}
\newcommand{\ket}[1]{\lvert#1\rangle}

\newcommand{\jump}[1]{\big[\hspace{-0.7mm} \big[ #1 \big]
  \hspace{-0.7mm} \big]} 
\newcommand{\mean}[1] {\big\{ \hspace{-0.7mm} \big\{ #1 \big\}
  \hspace{-0.7mm} \big\}}

\newcommand{\exc}{\epsilon_{\mathrm{xc}}}
\newcommand{\ext}{\mathrm{ext}}
\newcommand{\eff}{\mathrm{eff}}
\newcommand{\tot}{\mathrm{tot}}
\newcommand{\DG}{\mathrm{DG}}

\newcommand{\barint}{\kern4pt \raise3.4pt\hbox{\vrule height.6pt
    width7pt} \kern-11pt \int}

\journal{Journal of Computational Physics}

\begin{document}

\begin{frontmatter}



\title{Optimized local basis set for Kohn-Sham density
    functional theory}

    \author[ll]{Lin Lin\fnref{fnll}} 
\ead{linlin@lbl.gov}

\author[jl]{Jianfeng Lu} 
\ead{jianfeng@cims.nyu.edu}

\author[ly]{Lexing Ying}
\ead{lexing@math.utexas.edu}

\author[we]{Weinan E}
\ead{weinan@math.princeton.edu}

\address[ll]{Program in Applied and Computational Mathematics, Princeton
University, Princeton, NJ 08544, USA.}

\fntext[fnll]{Present affiliation: 
Computational Research Division, Lawrence
Berkeley National Laboratory, Berkeley, CA 94720, USA}

\address[jl]{Courant Institute of Mathematical
  Sciences, New York University, New York, NY 10012, USA.  }

\address[ly]{Department of Mathematics and ICES, University of Texas at
  Austin, Austin, TX 78712, USA. }

\address[we]{Department of Mathematics and PACM, Princeton University,
Princeton, NJ 08544, USA; Beijing International Center for Mathematical
Research, Peking University, Beijing, China 100871}

\begin{abstract}
 
  We develop a technique for generating a set of optimized local basis
  functions to solve models in the Kohn-Sham density functional theory
  for both insulating and metallic systems. The optimized local basis
  functions are obtained by solving a minimization problem in an
  admissible set determined by a large number of primitive basis
  functions.  Using the optimized local basis set, the electron energy
  and the atomic force can be calculated accurately with a small
  number of basis functions. The Pulay force is systematically
  controlled and is not required to be calculated, which makes the
  optimized local basis set an ideal tool for ab initio molecular
  dynamics and structure optimization. We also propose a
  preconditioned Newton-GMRES method to obtain the optimized local
  basis functions in practice.  The optimized local basis set is able
  to achieve high accuracy with a small number of basis functions per
  atom when applied to a one dimensional model problem.

\end{abstract}


\end{frontmatter}

%------------------------------------------------------------------------------------

\section{Introduction}\label{sec:intro}

% In many scientific problems, the quantities of interest can be
% expressed as a functional of a set of parameter-dependent functions.
% In order to evaluate the quantities of interest in an accurate and
% efficient way, it is desirable to expand these parameter-dependent
% functions with a small set of basis functions.  For example, in the
% electronic structure theory, the electron energy for a fixed atomic
% configuration is a functional of the electron wavefunctions, which are
% parameterized by the atomic positions. The dimension of the parameter
% space is proportional to the number of atoms in the system and can be
% tens of thousands in practice.  The electron wavefunctions can be
% expanded into a small number of atomic
% orbitals~\cite{SlaterKoster1954} where each atomic orbital is chosen
% adaptively to be centered around the position of one atom.  Similar
% idea of choosing parameter-dependent basis functions has also been
% explored in other contexts, such as the reduced basis function method
% for parameterized time harmonic Maxwell
% equations~\cite{ChenHesthavenMadayEtAl2010}, and the reduced-order
% model for the aeroelastic analysis~\cite{LieuFarhat2007}.  In this
% paper we focus on the Kohn-Sham density functional theory
% (KSDFT)~\cite{HohenbergKohn:64,KohnSham:65}, which is the most widely
% used electronic structure theory for condensed matter systems. The
% methods and concepts illustrated in this paper can be useful for other
% fields beyond KSDFT.

In scientific computation of systems with large number of degrees of
freedom, an efficient choice of basis functions becomes desirable
in order to reduce the computational cost. In this paper, we focus on the
choice of efficient basis sets for the Kohn-Sham density functional
theory (KSDFT)~\cite{HohenbergKohn:64,KohnSham:65}, which is the most
widely used electronic structure theory for condensed matter
systems. The methods and concepts illustrated here are also useful for
other applications.

In KSDFT, the quantities of interest are the electron energy $E(R)$
and the atomic force $F(R)$.  Here we denote by $R=(R_1, R_2, \cdots,
R_{N_A})^T$ the atomic positions, where $N_A$ is the number of atoms.
The atomic force is expressed in terms of the derivatives of the
electron energy with respect to the atomic positions as
$F(R)=-\frac{\partial E(R)}{\partial R}$. This is an important
quantity in many applications including structure optimization and
first principle molecular dynamics. The electron energy is a
functional of a set of Kohn-Sham orbitals $\{\psi_i\}_{i=1}^{N}$ where
$N$ is the number of electrons in the system. To illustrate the idea
with minimal technicality, let us consider for the moment a system of
non-interacting electrons at zero temperature.
% The full Kohn-Sham density functional theory will be introduced in
% Section~\ref{sec:dg}.
The energy functional for non-interacting electrons takes the form
\begin{equation}
  E(\{\psi_{i}(x)\}_{i=1}^{N};R)
  = \frac12 \sum_{i=1}^{N}\int \abs{\nabla
    \psi_i(x)}^2\ud x + \int  V(x;R)
  \sum_{i=1}^{N} \abs{\psi_{i}(x)}^{2}\ud x.
  \label{eqn:Esimplify}
\end{equation}
The first term and the second term in~\eqref{eqn:Esimplify} are
the kinetic energy and the potential energy of the system,
respectively. The energy $E(R)$ as a function of atomic positions is
given by the following minimization problem
\begin{equation}
  \begin{split}
    & E(R) = \min_{\{\psi_{i}(x)\}_{i=1}^{N}}
    E(\{\psi_{i}(x)\}_{i=1}^{N};R), \\
    &\text{s.t.} \quad \int \psi_{i}^{*}(x) \psi_{j}(x)\ud x = \delta_{ij},
    \quad i,j=1,\ldots,N.
  \end{split}
  \label{eqn:Eminpsi}
\end{equation}
We denote by $\{\psi_{i}(x;R)\}_{i=1}^{N}$ the minimizer. It can be
readily shown that $\{\psi_{i}(x;R)\}_{i=1}^{N}$ are the lowest $N$
eigenfunctions of the Hamiltonian operator $H(R)$, which takes the
form
\begin{equation}
  H(R) = -\frac12 \Delta_x + V(x;R).
  \label{eqn:Hsimplify}
\end{equation}
Using the Hamiltonian operator, the electron energy has an alternative
expression without the explicit dependence on the orbitals
$\{\psi_i\}_{i=1}^{N}$:
\begin{equation}
  E(R) = \Tr\left[ H(R) \chi(H(R)-\mu(R)) \right] \equiv \Tr[g_0(H(R))],
  \label{eqn:Ematrix}
\end{equation}
where $\chi(\cdot)$ is the Heaviside function: $\chi(x) = 1$ if $x<0$
and is $0$ otherwise. Here $\mu(R)$ is the chemical potential, which
takes value between the $N$-th and $(N+1)$-th eigenvalues of $H$ to
control the number of electrons.
%The second equality of Eq.~\eqref{eqn:Esimplify}
%defines a function $g(H)$ which will be extensively used in this
%paper.  
%The force can be expressed similarly as
%\begin{equation}\label{eqn:Fsimplify}
%    F = - \frac{\partial E}{\partial R} = - \Tr \left[ g'(H) 
%    \frac{\partial H}{\partial R} \right].
%\end{equation}

Since all the quantities depend on the atomic positions $R$, to
simplify the notation we drop the dependence of $R$ unless otherwise
specified.  If we approximate the eigenfunctions
$\{\psi_{i}\}_{i=1}^{N}$ by linear combination of a set of basis
functions $\Phi=(\phi_1,\cdots,\phi_{N_b})$, the Hamiltonian operator
$H$ is discretized into a finite dimensional matrix $\Phi^{\TT} H
\Phi$ (here and in the following, we will use the linear algebra
notation: $\phi_i^{\TT} H \phi_j = \averageop{\phi_i}{H}{\phi_j}$).
The number of basis functions $N_b$ is therefore called the
\textit{discretization cost}.  The electron energy and the force can
be expressed in terms of the discretized Hamiltonian operator as
\begin{equation}\label{eq:discE}
  \begin{split}
  E_{\Phi} & = \Tr \left[ g_0(\Phi^{\TT} H \Phi )\right],\\
  F_{\Phi,I} & = -\frac{\partial E_{\Phi}}{\partial
  R_I} \\
  &= - \Tr \left[ g_0'(\Phi^{\TT} H \Phi) \Phi^{\TT}\frac{\partial
  H}{\partial R_I} \Phi
  \right] - 2 \Tr \left[ g_0'(\Phi^{\TT} H \Phi) 
  \Phi^{\TT} H \frac{\partial \Phi}{\partial R_I} \right].
  \end{split}
\end{equation}
$F_{\Phi,I}$ is the $I$-th component of the
force. In what follows the second equation in \eqref{eq:discE} is also
written in a compact form as
\begin{equation}
  \begin{split}
  F_{\Phi} & = -\frac{\partial E_{\Phi}}{\partial
  R} \\
  &= - \Tr \left[ g_0'(\Phi^{\TT} H \Phi) \Phi^{\TT}\frac{\partial
  H}{\partial R} \Phi
  \right] - 2 \Tr \left[ g_0'(\Phi^{\TT} H \Phi) 
  \Phi^{\TT} H \frac{\partial \Phi}{\partial R} \right].
  \end{split}
  \label{}
\end{equation}

Choosing basis functions $\Phi$ adaptively with respect to the atomic
positions $R$ has obvious computational advantages, as it allows the
possibility to reduce the discretization cost by a significant amount
while maintaining the accuracy for the evaluation of the electron
energy and atomic forces.  Since the electron energy is defined
variationally as in~\eqref{eqn:Eminpsi}, an accurate basis set should
minimize the electron energy.  However, choosing the basis functions
adaptively gives arise to some difficulties in the evaluation of the
force~\eqref{eq:discE} which requires the calculation of
$\frac{\partial \Phi}{\partial R}$.  In electronic structure theory,
the contribution from $\frac{\partial \Phi}{\partial R}$ is referred
to as the Pulay force~\cite{Pulay:69}.  We will henceforth adopt this
terminology.  The Pulay force originates from the incompleteness of
the basis set, and has been found to be important to obtain the force
with reliable accuracy for structure optimization or first principle
molecular dynamics~\cite{Pulay:69,BendtZunger1983}.  The calculation
of the Pulay force can be quite expensive even if the basis functions
$\Phi$ have analytical expressions, and the calculation of the Pulay
force becomes almost intractable if the basis functions are defined
implicitly such as in the adaptive mesh
method~\cite{TsuchidaTsukada1996,GygiGalli1995,BylaskaHolstWeare2009,ZhangShenZhouEtAl2008}. We
would like to systematically reduce the Pulay force so that the
approximation
\begin{equation}
  \frac{\partial E_{\Phi}}{\partial R} 
  \approx 
  \Tr \left[ g_0'(\Phi^{\TT} H \Phi) \Phi^{\TT}\frac{\partial H}{\partial R} \Phi
  \right]
\end{equation}
becomes adequate.

The key observation in this paper is that minimizing the electron
energy and reducing the Pulay force can be simultaneously achieved by
the following optimization procedure
\begin{equation}
  \min_{\Phi\subset \mc{V}, \Phi^T\Phi=I} E_{\Phi} = \min_{\Phi\subset \mc{V}, \Phi^T\Phi=I}
  \Tr \left[ g_0(\Phi^{\TT} H \Phi )\right]
  \label{eqn:minE}
\end{equation}
Here $\mc{V}$ is an admissible subset of the space spanned by a set of
\textit{primitive basis functions} which are independent of $R$. Later
$\mc{V}$ will be referred to as the \textit{admissible set}.  We
select from $\mc{V}$ a small number of $R$-dependent \textit{optimized
  basis functions} $\Phi=(\phi_1,\cdots,\phi_{N_b})$ which give rise
to the lowest electron energy in $\mc{V}$. The Euler-Lagrange equation
for the minimization problem~\eqref{eqn:minE} reads
\begin{equation}\label{eq:firstopt}
  \begin{cases}
    H \Phi g_0'(\Phi^{\TT} H \Phi) = \Phi \Lambda \\
    \Phi^{\TT} \Phi = I
  \end{cases},
\end{equation}
where the matrix $\Lambda$ is a Lagrangian multiplier and is symmetric.  
When the first optimality condition \eqref{eq:firstopt} is
satisfied, we find
\begin{equation}\label{eqn:pulayvanish}
  2 \Tr \left[ g_0'(\Phi^{\TT} H \Phi) 
  \Phi^{\TT} H \frac{\partial \Phi}{\partial R} \right] = 
  2\Tr\left[ \Lambda \Phi^{\TT} \frac{\partial \Phi}{\partial R}\right]
  = \Tr\left[\Lambda \frac{\partial(\Phi^{\TT} \Phi)}{\partial R}
  \right] = 0.
\end{equation}
The last equality comes from the orthonormal constraint on the
optimized basis functions $\Phi$.  The reason
why~\eqref{eqn:pulayvanish} holds can be understood from the
variational structure of the original problem~\eqref{eqn:minE}, which
is related to the Hellmann-Feynman theorem in quantum mechanics. As a
result, the Pulay force vanishes in the atomic force even if the
optimized basis functions are far from being a complete basis set.

The choice of the primitive basis functions is crucial.  Although the
optimized basis functions are always incomplete due to the small
number of basis functions used, the primitive basis set should be
systematically improvable towards a complete basis set. Each primitive
basis function should be local in order to be suitable for large scale
parallel calculation. In our previous
work~\cite{LinLuYingE_Adaptive}, the primitive basis set is
constructed using a discontinuous Galerkin (DG) framework. The DG
primitive basis set allows the usage of basis functions that are
discontinuous across element surfaces. Each DG primitive basis
function is local in the real space, and thus gives full flexibility
in the choice of the optimized basis functions.  The locality
constraint in the real space can therefore be naturally applied to the
optimized basis functions, giving rise to the \textit{optimized local
  basis set}.

We remark that a large primitive basis set also presents practical
difficulties for the optimization procedure.  In this paper we propose
a preconditioned Newton-GMRES method to obtain the optimized local
basis functions.  Numerical results using a one dimensional model
problem validate the performance of the optimized local basis
functions: the electron energy and the force can be accurately
calculated along the trajectory of the molecular dynamics without
systematic drift, using a very small number of basis functions per
atom.

Improving the quality of the basis functions via variational
optimization has been previously studied in the electronic structure
theory. However, to the best of our knowledge all the optimized basis
functions presented so far use atom-centered primitive basis
functions, such as atomic orbitals or Gaussian-type orbitals.  Since
atomic orbitals or Gaussian-type orbitals depend on the atomic
positions and do not form a complete basis set, the Pulay force never
vanishes. The Pulay force of all the primitive basis functions should
be computed for each atomic configuration.  Moreover, optimization for
each atomic configuration is generally considered to be an expensive
procedure, and the optimized basis functions are usually obtained for
specific reference systems instead. For example, Junquera
\etal~\cite{JunqueraSIESTA2001} proposed to optimize the shape and
cutoff radii of a set of numerical atomic orbitals;
Ozaki~\cite{Ozaki2003} proposed using the optimal linear combination
of a set of numerical atomic orbitals; Blum
\etal~\cite{BlumGehrkeHankeEtAl2009} used a greedy method to select
basis functions from a large pool of numerical atomic orbitals.  The
drawback of this procedure is that the quality of the basis functions
depends heavily on the choice of the reference system.  The
transferability of these basis sets obtained for specific reference
systems should be tested carefully for a variety of systems.
% , and this issue resembles that in the pseudopotential
% community~\cite{TroullierMartins1991}.
Optimized basis functions without the choice of reference systems have
also been studied before.  Talman~\cite{Talman2000} proposed to
optimize a set of numerical atomic orbitals for all the atoms
simultaneously.  Rayson and Briddon~\cite{RaysonBriddon2009} tried to
find the optimal linear combination of Gaussian-type orbitals, where
the optimization process loops over each atom in the system.  These
methods share similar spirit as the present work, and can be regarded
as approximate strategies towards achieving optimality in practice.

Our current work avoids the subtle issue of transferability by
means of an optimization procedure for any given system, which could be
advantageous for complex systems where manually constructed transferable
basis functions are difficult to be obtained.  The DG primitive basis
set is a complete basis set, and the optimized local basis functions are
local by construction.  The DG primitive basis set is independent of the
atomic positions, and the Pulay force vanishes when the optimality
condition is reached.

The rest of the paper is organized as follows.  In
Section~\ref{sec:opt}, we introduce the optimized local basis set for
KSDFT.  Numerical examples are presented in Section~\ref{sec:numer},
followed by discussion and conclusion in Section~\ref{sec:conclusion}.
To make the paper self-contained, we briefly recall the finite
temperature Kohn-Sham density functional theory in \ref{sec:dg}.

\section{Optimized local basis function} \label{sec:opt}

As introduced in our previous work \cite{LinLuYingE_Adaptive}, using a
discontinuous Galerkin method (the interior penalty method
\cite{BabuskaZlamal:73, Arnold:82}), the effective energy functional
in Kohn-Sham density functional theory is given by
\begin{equation}\label{eq:DGvar}
  \begin{aligned}
    \mc{F}_{\DG}(\{\psi_i\}, \{f_i\}) & = \frac{1}{2} \sum_{i} f_i
    \average{\nabla \psi_i , \nabla \psi_i}_{\mc{T}} - \sum_{i} f_i
    \average{\mean{\nabla\psi_i}, \jump{\psi_i}}_{\mc{S}}
    + \average{ V_{\eff}, \rho }_{\mc{T}} \\
    & + \alpha \sum_{i} f_i \average{\jump{\psi_i},
      \jump{\psi_i}}_{\mc{S}} + \sum_{\ell} \gamma_{\ell} \sum_{i} f_i
    \abs{\average{b_{\ell}, \psi_i}_{\mc{T}}}^2 \\
    & + \beta^{-1} \sum_i \bigl( f_i \ln f_i + (1-f_i) \ln (1 - f_i)
    \bigr).
  \end{aligned}
\end{equation}
This is a discretization method for the Helmholtz free energy
\eqref{eq:linearvar} for a system at temperature $\beta^{-1}$, see
\ref{sec:dg} for details of formulation of Kohn-Sham density
functional theory in finite temperature. Here $\mc{T}$ is a collection
of quasi-uniform rectangular partitions of the computational domain:
\begin{equation}
  \mc{T} = \{E_1, E_2, \cdots, E_M \},
\end{equation}
and $\mc{S}$ be the collection of surfaces that correspond to
$\mc{T}$. $\average{\cdot, \cdot}_{\mc{T}}$ and $\average{\cdot,
  \cdot}_{\mc{S}}$ are inner products in the bulk and on the surface
respectively. The notations $\mean{\cdot}$ and $\jump{\cdot}$ are used
for the standard average and jump operators across surfaces in the
interior penalty method. We refer to \cite{LinLuYingE_Adaptive} for
more details.

Let $\Phi$ be a chosen set of basis functions $\Phi =
\{\varphi_{k,j}\}_{j=1}^{J_k}$, where each $\varphi_{k,j}$ is
supported in $E_k$ and $J_k$ is the total number of basis functions in
$E_k$.  The corresponding approximation space $\mc{V}_{\Phi}$ is given
by
\begin{equation}
  \mc{V}_{\Phi} = \spanop\{ \varphi_{k,j},\, E_k \in \mc{T},\, 
  j = 1, \cdots, J_k \}.
\end{equation}
The approximated Kohn-Sham orbitals are the solutions
to the minimization problem
\begin{equation}\label{eqn:KSfunc_DG}
  \begin{split}
    &\min_{\{\psi_{i}\}\subset \mc{V}_{\Phi},\{f_i\}} \mc{F}_{\DG}(\{\psi_i\},\{f_i\}),\\
    &\text{s.t.} \quad \int \psi_i^{\ast} \psi_j \ud x =
    \delta_{ij},\quad i,j=1,\cdots,\widetilde{N},
  \end{split}
\end{equation}
where $\widetilde{N}$ is chosen to be slightly larger than the number
of electrons $N$ in the system in order to compensate for the finite
temperature effect (see \ref{sec:dg} for more detailed explanation).
We propose the optimized local basis functions which give rise to a
specific choice of $\Phi$, in order to achieve accuracy for both the
Helmholtz free energy and the force while using a small number of
basis functions. Following the spirit of~\eqref{eqn:minE} introduced
for the model problem in the introduction, the optimized local basis
function set $\Phi$ solves the following minimization problem
\begin{equation}\label{eq:doublemin}
  \min_{\Phi\subset \mc{V}, \Phi^T\Phi=I} \min_{\{\psi_i\} \subset \mc{V}_{\Phi}, \{f_i\}} 
  \mc{F}_{\DG}(\{\psi_i\}, \{f_i\}), 
\end{equation}
where $\mc{V}$ is the \textit{admissible set}.
%Given the
%effective Hamiltonian $H$ (we drop the subscript $\mathrm{eff}$ for
%simplicity from now on) and a basis set $\Phi$, we minimize
%\eqref{eq:DGvar} for $\{ \psi_i\} \subset \mc{V}_{\Phi}$. We want to
%choose a good basis set $\Phi$ so that the number of basis functions
%is small while the accuracy is guaranteed. Therefore, we are going to
%optimize over different choices of $\Phi$:
%so that the resulting $\psi_i$'s and hence the density $\wt{\rho}$
%will be close to the true results. 
%To solve the optimization problem \eqref{eq:doublemin}, we have to
%discretize the problem first. For this, we will choose an underlying
%basis sets $U$. It should not be confused with the desired basis sets
%$\Phi$ which is the solution of \eqref{eq:doublemin}. We will
%represent the Hamiltonian and also the desired basis functions $\Phi$
%in the basis set $U$.
To define the admissible set, we take for each element $E_k$ a set of
basis functions $\{u_{k, j}, j = 1, \cdots, J_k\}$.  Each $u_{k,j}$ is
compactly supported in $E_k$, and they satisfy the orthonormality
condition
\begin{equation}
  \average{u_{k',j'}, u_{k, j}}_{\mc{T}} = \delta_{kk'} \delta_{jj'}.
\end{equation}
For example, $\{u_{k,j}\}$ can be polynomials restricted to the set
$E_k$ up to a certain order.  Other forms of primitive basis functions
can be chosen as well, without changing the discussion that follows.
The discretized Hamiltonian in the DG formulation takes the form
\begin{equation}\label{eq:discH}
  \begin{aligned}
    H_{k',j';k,j} = & \frac{1}{2}\average{\nabla u_{k',j'}, \nabla
      u_{k,j}}_{\mc{T}} -\frac{1}{2}\average{\jump{ u_{k',j'}},
      \mean{\nabla u_{k,j}}}_{\mc{S}} \\
    & -\frac{1}{2}\average{\mean{\nabla u_{k',j'}},
      \jump{u_{k,j}}}_{\mc{S}} + \alpha
    \average{\jump{u_{k',j'}},
      \jump{u_{k,j}}}_{\mc{S}} \\
    & +\average{u_{k',j'}, V_{\eff} u_{k,j}}_{\mc{T}} + \sum_\ell
    \gamma_{\ell} \average{u_{k',j'},b_\ell}_{\mc{T}} \average{b_\ell,
      u_{k,j}}_{\mc{T}}.
  \end{aligned}
\end{equation}

The optimized local basis functions should be local in the real space
in order to facilitate large scale computation.  Since $\{u_{k,j}\}$
are compactly supported in $E_k$, the locality constraint on the
optimized local basis functions is naturally imposed by requiring each
function in the admissible set to be linear combinations of
$\{u_{k,j}\}$ for the same $k$, \ie
\begin{equation}
  \mc{V}=\bigcup_{k=1}^{M} \spanop\{ u_{k,j}, j = 1, \cdots, J_k \},
  \label{eqn:primitive}
\end{equation}
where $M$ is the number of elements.

Inside each element $E_k$, we select $N_k$ optimized local basis
functions from the admissible set.  $N_k$ is much smaller than $J_k$.
The optimized local basis functions are denoted by $\{\phi_{k,1},
\cdots, \phi_{k,N_k} \}$, and are represented by the linear
combination of the primitive basis functions
\begin{equation*}
  \phi_{k,l} = \sum_{j=1}^{J_k} \wt{\phi}_{k,l,j} u_{k,j}, \quad l = 1, \cdots, N_k.
\end{equation*}
With slight abuse of notation, we use $\phi_{k,l}$ also for the column
vector of the coefficients in the primitive basis functions:
\begin{equation}
  \phi_{k,l} = 
  \begin{pmatrix} 
    \wt{\phi}_{k,l,1} & \wt{\phi}_{k,l,2} & \cdots & \wt{\phi}_{k,l, J_K} 
  \end{pmatrix}^{\TT}.
\end{equation}
If we write
\begin{equation}
  \Phi_k = 
  \begin{pmatrix}
    \phi_{k, 1} & \phi_{k,2} & \cdots & \phi_{k,N_k}
  \end{pmatrix},
\end{equation}
the optimized local basis set $\Phi$ represented in the primitive basis set
takes the form
% \begin{equation}
%   \Phi = 
%   \begin{pmatrix}
%     \Phi_1 \\
%     & \Phi_2 \\
%     & & \ddots \\
%     & & & \Phi_M
%   \end{pmatrix}.
% \end{equation}
\begin{equation}
  \Phi = \diag( \Phi_1, \Phi_2, \cdots, \Phi_M).
\end{equation}
Because of the block diagonal
structure, the orthonormality constraint $\Phi^T\Phi=I$ is equivalent to
the orthonormal constraint for each $\Phi_{k}$, \ie,
$\Phi_{k}^T\Phi_k=I_k,k=1,\cdots,M$.  Here each block $\Phi_{i}$ is a
rectangular matrix of size $N_{g}\times N_k$, where $N_{g}$ is the
number of grid points in the element, and $N_k$ is the number of basis
functions. $I_k$ is an $N_k \times
N_k$ identity matrix.

Under the basis set $\Phi$, the discretized Hamiltonian becomes
$\Phi^{\TT} H \Phi$ with $H$ given by \eqref{eq:discH}. The Helmholtz
free energy can be written without the explicit dependence on
$\{\psi_{i}\}$ and $\{f_i\}$:
\begin{equation}
 \min_{\{\psi_i\} \subset \mc{V}_{\Phi}, \{f_i\}} 
  \mc{F}_{\DG}(\{\psi_i\}, \{f_i\})  = \Tr g(\Phi^{\TT} H\Phi) + \mu N,
\end{equation}
where the function $g$, which is a finite temperature version of
$g_0$, is defined as
\begin{equation}\label{eqn:gfunc}
  g(x) = - \beta^{-1} \ln ( 1 + \exp(\beta( \mu - x)) ).
\end{equation}
Note that the derivative of $g$ is the Fermi-Dirac function
\begin{equation}
  g'(x) = ( 1 + \exp(\beta(x - \mu)) )^{-1}.
\end{equation}
Hence, the minimization problem~\eqref{eq:doublemin} becomes 
\begin{equation}\label{eq:canonicalenergy}
  \begin{split}
  &\mc{F}_{\DG} = \min_{\Phi\subset \mc{V}} \left[ \Tr
  g(\Phi^{\TT} H\Phi) + \mu N \right],\\
  & \text{s.t.} \quad \Phi_k^T\Phi_k = I_k,\quad k=1,\cdots,M.
  \end{split}
\end{equation}
The atomic force is then given by 
\begin{equation}
  \begin{aligned}
    F & = - \frac{\partial \mc{F}_{\DG}}{\partial R} \\
    & = - \Tr( \rho_{\Phi} \Phi^{\TT} \frac{\partial H}{\partial R} \Phi) - 2
    \Tr ( \rho_{\Phi} \Phi^{\TT} H \frac{\partial \Phi}{\partial R})\\
    & = - \Tr( \rho_{\Phi} \Phi^{\TT} \frac{\partial H}{\partial R}
    \Phi),
  \end{aligned}
\end{equation}
where $\rho_{\Phi}= g'(\Phi^{\TT} H \Phi)$ is the single particle
density matrix associated to the discretized Hamiltonian $\Phi^{\TT} H
\Phi$. $\rho_{\Phi}$ can be evaluated using standard
diagonalization techniques by computing the 
eigenvalues and eigenvectors of the reduced Hamiltonian $\Phi^T H\Phi$.  
This is asymptotically the most time consuming step which scales as
$O(N^3)$ where $N$ is the number of electrons in the system. For the 1D
system considered in this manuscript, $\rho_{\Phi}$ is solved by the
MATLAB diagonalization subroutine \textit{eig}. For systems of large
size, the diagonalization routine can be replaced by the recently
developed low order scaling selected inversion
methods~\cite{LinLuYingE2009,LinYangMezaEtAl2010} to reduce the
computational cost.  The Pulay force vanishes in the last equality when
the first order optimality of the optimization problem
\eqref{eq:canonicalenergy} is reached, following the same reasoning as
in~\eqref{eqn:pulayvanish}.

%The gradient of $\Omega_{\Phi}$ with respect to $\Phi$ is given by
%\begin{equation}
%  \frac{\partial \Omega_{\Phi}}{\partial \Phi} = 2 \rho_{\Phi} \Phi^{\TT} H.
%\end{equation}
%We observe that the Pulay force can be written as 
%\begin{equation}
%  - 2 \Tr ( \rho_{\Phi} \Phi^{\TT} H \frac{\partial \Phi}{\partial R_I})
%  = - \Tr ( \frac{\partial \Omega_{\Phi}}{\partial \Phi} \frac{\partial \Phi}
%  {\partial R_I}),
%\end{equation}
%hence the first-order optimality condition of \eqref{eq:optomega}
%guarantees that the Pulay force vanishes. 

The Euler-Lagrange equation with respect to the minimization
problem~\eqref{eq:canonicalenergy} reads
\begin{equation}\label{eq:firstopt2}
  \begin{cases}
    H \Phi \rho_{\Phi} - \Phi \Lambda = 0 \\
    \Phi^{\TT} \Phi - I = 0
  \end{cases},
\end{equation}
where the $\Lambda$ is a block diagonal matrix
\begin{equation*}
  \Lambda = \diag(\Lambda_1, \Lambda_2, \cdots, \Lambda_M),
\end{equation*}
which is the Lagrange multiplier for the orthonormal constraints.
Due to the block diagonal structure of $\Phi$, we can write the first
order optimality condition~\eqref{eq:firstopt2} as
\begin{align}
  \sum_j H_{ij} \Phi_j \rho_{\Phi, ji} - \Phi_i \Lambda_i = 0, \quad i
  = 1, \cdots, M.
\end{align}
Define the remainder for the $i$-th element as
\begin{equation}
  R_i(\Phi, \Lambda) =
  \begin{pmatrix}
    \sum_j H_{ij} \Phi_j \rho_{\Phi, ji} - \Phi_i \Lambda_i \\
    I - \Phi_i^{\TT} \Phi_i 
  \end{pmatrix}.
\end{equation}
We solve $R_i(\Phi, \Lambda) = 0$ for $i = 1, 2, \cdots, M$.

In order to solve the nonlinear system~\eqref{eq:firstopt2},
we propose a preconditioned Newton-GMRES method as follows.  Denote
by $J$ the Jacobian matrix. At the $l$-th iteration, the Newton step
solves the following linear system for the correction term
\begin{equation}\label{eq:inneriter}
  J^{(l)} 
  \begin{pmatrix} 
    \Delta \Phi^{(l)} \\
    \Delta \Lambda^{(l)}
  \end{pmatrix} = - 
  \begin{pmatrix}
    H \Phi g'(\Phi^{\TT} H \Phi) - \Phi \Lambda \\
    I - \Phi^{\TT} \Phi 
  \end{pmatrix}.
\end{equation}
%After the correction direction is found by solving
%\eqref{eq:inneriter}, $\Phi$ and $\Lambda$ of the next iteration will
%be obtained using line search. 

To make the optimization feasible in practice, we take the following
approximation.  We neglect the derivative of $\rho_{\Phi} =
g'(\Phi^{\TT} H \Phi)$ with respect to $\Phi$ in the Jacobian. The
most important reason for this approximation is that the numerical
evaluation of such derivative is quite expensive.  In practice we find
that the residue of the Euler-Lagrange equation decays fast in the
first few Newton iterations, and slows down when the residue becomes
small, suggesting that the derivative of $\rho_{\Phi}$ with respect to
the basis functions can be important especially for the small residue
case.  Numerical results indicate that the accuracy of the Helmholtz
free energy and the force can already be improved by one order of
magnitude after a few Newton iterations.  Further improvement that
includes the approximate form of the derivative of $\rho_{\Phi}$ will
be considered in the future work.  Using this approximation, the
correction equation \eqref{eq:inneriter} can be written explicitly as
\begin{equation}\label{eq:jacobian}
  \begin{pmatrix}
    \sum_j H_{ij} (\Delta \Phi)_j \rho_{\Phi, ji} - (\Delta \Phi)_i \Lambda_i 
    - \Phi_i (\Delta \Lambda)_i \\
    - \Phi_i^{\TT} (\Delta \Phi)_i - (\Delta \Phi)^{\TT}_i \Phi_i 
  \end{pmatrix} = - R_i,
\end{equation}
for $i = 1, 2, \cdots, M$. 

We solve the linear system~\eqref{eq:jacobian} using a preconditioned
GMRES method.  The GMRES method~\cite{SaadSchultz1986} is a robust way for solving
ill-conditioned linear equations. The preconditioner should give an
approximate solution efficiently for the following equation 
\begin{equation}
  \begin{pmatrix}
    \sum_j H_{ij} (\Delta \Phi)_j \rho_{\Phi, ji} - (\Delta \Phi)_i \Lambda_i 
    - \Phi_i (\Delta \Lambda)_i \\
    - \Phi_i^{\TT} (\Delta \Phi)_i - (\Delta \Phi)^{\TT}_i \Phi_i 
  \end{pmatrix} = - 
  \begin{pmatrix}
    B_i \\
    C_i 
  \end{pmatrix}
  \label{eqn:precond}
\end{equation}
for any right hand side $\{B_i\},\{C_i\}$. To this end we first
neglect the interaction between different elements:
\begin{equation}\label{eq:diagjacob}
  \begin{pmatrix}
    H_{ii} (\Delta \Phi)_i \rho_{\Phi, ii} - (\Delta \Phi)_i \Lambda_i
    - \Phi_i (\Delta \Lambda)_i \\
    - \Phi_i^{\TT} (\Delta \Phi)_i - (\Delta \Phi)^{\TT}_i \Phi_i
  \end{pmatrix} = -
  \begin{pmatrix}
    B_i \\
    C_i 
  \end{pmatrix}.
\end{equation}
The equations of $(\Delta \Phi)_i$ for different elements become
decoupled. \eqref{eq:diagjacob} can be therefore solved independently
in each element. Second, we note that there are degeneracy issues
solving \eqref{eq:diagjacob}. This is because in the subspace spanned
by the basis $\mc{V}_{\Phi}$, only the low-lying eigenfunctions of the
discrete Hamiltonian affect the free energy much, while the
eigenfunctions with large eigenvalues do not contribute much due to
small occupation number. Therefore, if we change the subspace
$\mc{V}_{\Phi}$ in the direction of these high energy eigenfunctions,
it does not change much the energy, which causes degeneracy.

We propose the following pruning method to solve the degeneracy
problem.  Instead of solving \eqref{eq:diagjacob}, we restrict to the
basis functions contributed to the low-lying eigenfunctions by the
following procedure. Given density matrix $\rho_{\Phi}$, for each
element $E_i$, we take a singular value decomposition of the diagonal
block of $\rho_{\Phi, ii}$:
\begin{equation}
  \rho_{\Phi, ii} = U_i S_i U_i^{\TT},
\end{equation}
with the singular values sorted in descending order. Then according to
magnitude of the singular values, we write $U_i = ( U_i^h, U_i^l)$,
where the singular vectors in $U_i^h$ correspond to high singular
values above a certain threshold, and the ones in $U_{i}^{l}$
correspond to low singular values below the threshold. The basis
functions in the element can be separated into two accordingly:
\begin{equation}
  \Phi_i^h = \Phi_i U_i^h, \qquad \Phi_i^l = \Phi_i U_i^l.
\end{equation}
We now only update the correction term corresponding to the high singular values by solving
\begin{equation}\label{eq:diagjacobs}
  \begin{pmatrix}
    H_{ii} (\Delta \Phi)_i^h \rho_{\Phi, ii}^h - (\Delta \Phi)_i^h \Lambda_i^h
    - \Phi_i^h (\Delta \Lambda)_i^h \\
    - (\Phi_i^h)^{\TT} (\Delta \Phi^h)_i - (\Delta \Phi^h)^{\TT}_i \Phi_i^h
  \end{pmatrix} = - 
  \begin{pmatrix}
    B_i U_i^h \\
    (U_i^h)^T C_i U_i^{h}
  \end{pmatrix},
\end{equation}
where 
\begin{equation*}
  \rho_{\Phi, ij}^h = (U_i^h)^{\TT} \rho_{\Phi, ij} U_j^h.
\end{equation*}
The approximate solution of the preconditioning
equation~\eqref{eqn:precond} is therefore given by
\begin{equation}
 \Delta \Phi_i = \Delta \Phi_i^h (U_i^h)^T, 
 \Delta \Lambda_i = U_i^h \Delta \Lambda_{i}^{h} (U_i^h)^T.
  \label{}
\end{equation}

As will be seen in the numerical examples in Section~\ref{sec:numer},
the preconditioned Newton-GMRES method is able to obtain the optimized
local basis functions efficiently with a small number of iterations.

\section{Numerical result}
\label{sec:numer}

\subsection{Setup}

The accuracy and efficiency of the optimized local basis functions is
illustrated using a one-dimensional model problem as follows.  The
number of atoms in the one-dimensional model problem is denoted by
$N_A$, the positions of electrons by $x$, and the positions of ions by
$R=(R_1, R_2, \cdots, R_{N_A})^T$.  The electronic and ionic degrees
of freedom are separated by the Born-Oppenheimer approximation. The
effective Kohn-Sham Hamiltonian of the electrons for a given atomic
configuration $R$ is
\begin{equation}
  H(R) = -\frac12 \Delta + V(x;R).
  \label{}
\end{equation}
The effective electron-ion interaction and electron-electron
interaction is modeled by the summation of a series of Gaussian
functions
\begin{equation}
  V(x;R) = -\frac{A}{\sqrt{2\pi\sigma^2}} \sum_{I=1}^{N_A} e^{-
  \frac{(x-R_I)^2}{2\sigma^2}}.
  \label{eqn:pot}
\end{equation}
$A$ and $\sigma$ characterize the height and the width of the
potential well around each atom, respectively. For simplicity, the
effective Hamiltonian does not depend on the electron density, and
hence self-consistency iteration is not involved. The self-consistent
iteration will be added
in the future work. The ion-ion interaction is modeled by a harmonic
potential with periodized nearest-neighbor interaction
\begin{equation}
  V_{II}(R) = \frac12 \sum_{I=1}^{N_A-1} \omega (R_I - R_{I+1})^2 +
  \frac12\omega (R_{N_A}-R_{1}-L)^2,
  \label{}
\end{equation}
with $L$ being the length of the computational domain.
The force on atom $I$ is 
\begin{equation}
  F_I = -\frac{\partial \mc{F}_{\DG}(R)}{\partial R_I} -\frac{\partial
  V_{II}(R)}{\partial R_I}.
  \label{eqn:HF1D}
\end{equation}
The finite temperature KSDFT is used here and the
Helmholtz free energy for the electrons
$\mc{F}_{\DG}(R)$ is given by \eqref{eq:canonicalenergy}. 
The finite temperature effect is usually negligible in
insulating systems with large band gap, but becomes important for the
stability in metallic systems with small or vanishing band gap.  

The accuracy is measured in terms of the error of the Helmholtz free
energy per atom and the error of the force.  For a given atomic
configuration, the Helmholtz free energy per atom and the force are
calculated independently using the optimized local basis functions and
the benchmark plane wave basis functions.  Except for the unit of
temperature which is Kelvin, atomic units are used throughout this
section unless otherwise specified.  In particular, the unit of energy
is Hartree, the unit of force is Hartree/Bohr, and the electron mass
$m$, electron charge $e$ and the Planck constant $\hbar$ are set to be
unity.  The detailed choices of the parameters in the simulation are
as follows. Except in the last example where we test for different
system sizes, the number of atom is taken to be $N_A = 8$. The average
distance between adjacent atoms is $10$ au, and the size of each
element is also set to be $10$ au.  The initial guess of the optimized
local basis functions uses the adaptive local basis functions proposed
in our previous work~\cite{LinLuYingE_Adaptive}. The adaptive local
basis functions use a small buffer region outside each element. The
buffer size is $5$ au in the present calculation. We compare the
electron energy and the forces produced by the optimized local basis
functions with those obtained from a planewave calculation with
kinetic energy cutoff at $E_{\textrm{cut}}=40$ Ry, or $20$ planewaves
per atom. The change of the Helmholtz free energy and the force is
less than $10^{-8}$ au if the kinetic energy cutoff for the planewave
calculation is further increased.  $21$ Legendre-Gauss-Lobatto (LGL)
grid points per element are used to discretize the optimized local
basis functions as well as the adaptive local basis functions.  The
change of the Helmholtz free energy and the force is less than
$10^{-8}$ au if the number of LGL integration points is further
increased. Therefore the numerical integration error is negligible,
and the error in the calculated Helmholtz free energy and the force
faithfully represents the error due to the usage of adaptive local
basis functions or optimized local basis functions.  The electron
temperature is $2000$ K.  The penalty parameter $\alpha$ in the DG
Hamiltonian is $40$.  The choice of parameters for the potential
energy surface is $\omega=0.03, A=5.0, \sigma=4.0$.

If one electron is assigned to each atom (spin
degeneracy is neglected), then the band gap at the equidistant
configuration is around $14000$ K, which is much larger than the
electron temperature ($2000$ K).  In what follows this system is
referred to as the insulating system. If four electrons are assigned to
each atom, the band gap is is essentially zero ($0.5$ K). The energy
levels around the Fermi surface are fractionally occupied due to the
thermal effect.  This system is referred to as the metallic system. 

In the optimization of the local basis functions, the maximum number
of Newton iterations is set to be $4$, and the maximum number of
iterations for the preconditioned GMRES solver for the Newton's
equation is set to be $30$. We find that the error for solving the
linear system~\eqref{eq:jacobian} using $30$ preconditioned GMRES
iterations is less than $10^{-4}$.  The threshold value for the
significant part of the basis functions is set to be $10^{-7}$ to
avoid degeneracy.  The preconditioning step is solved by direct $LU$
decomposition method inside each element.

\subsection{Static case}\label{subsec:static}

We first illustrate the performance of the optimized local basis set
in the static case. $20$ atomic configurations are generated from
equidistant configuration with small random perturbations.  The
accuracy of using the optimized local basis set is measured by the
mean absolute value of the error (mean error) of the Helmholtz free
energy per atom and the mean error of the force of a fixed atom.
Besides the optimized local basis functions, the error of using the
adaptive local basis functions~\cite{LinLuYingE_Adaptive} is presented
as well to illustrate the effectiveness of the optimization procedure.

\begin{figure}[h]
  \begin{center}
    \subfloat[$\phi_{1}$]{\includegraphics[width=0.35\textwidth,
    height=0.35\textwidth]{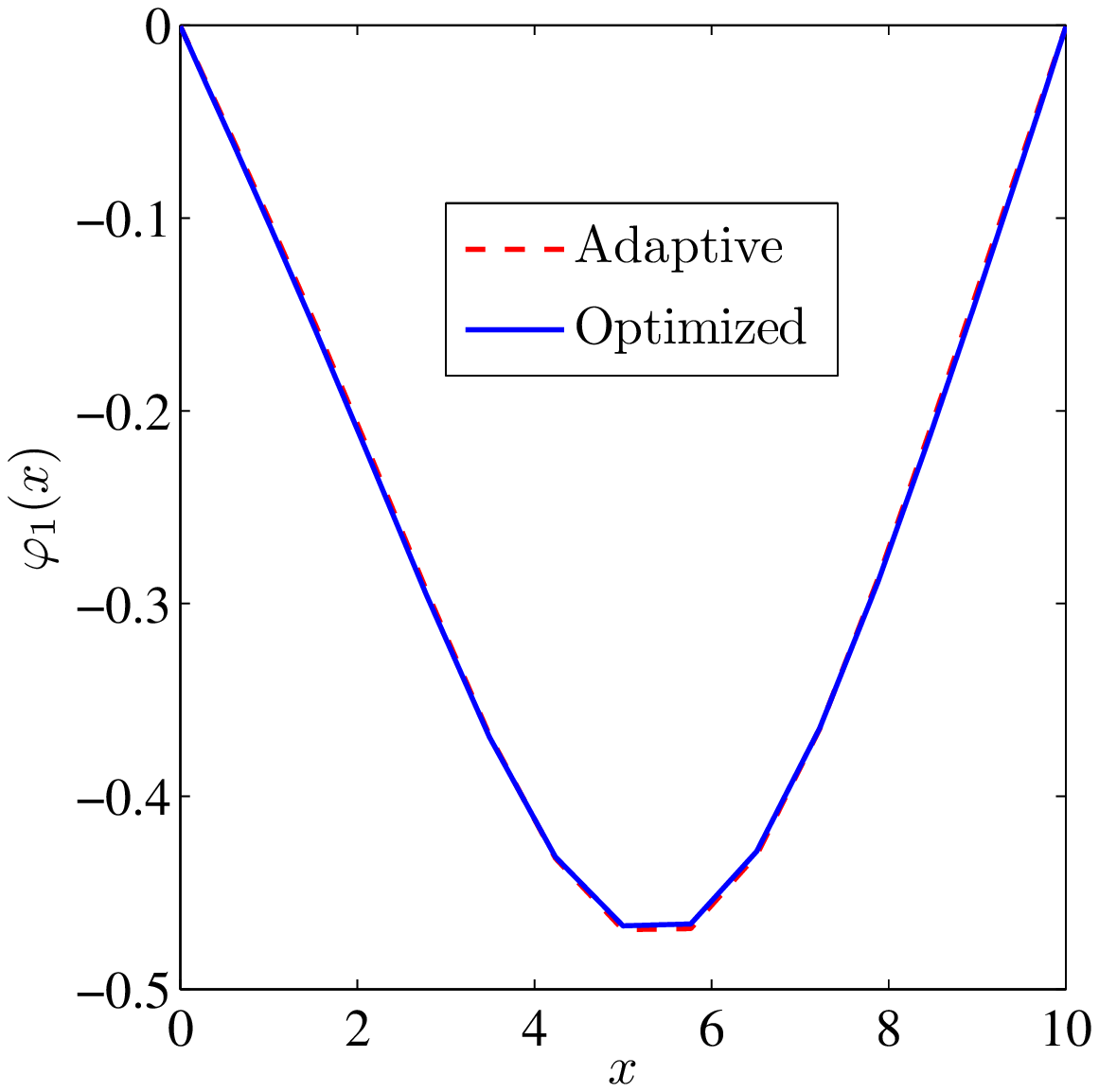}}
    \subfloat[$\phi_{2}$]{\includegraphics[width=0.35\textwidth,
    height=0.35\textwidth]{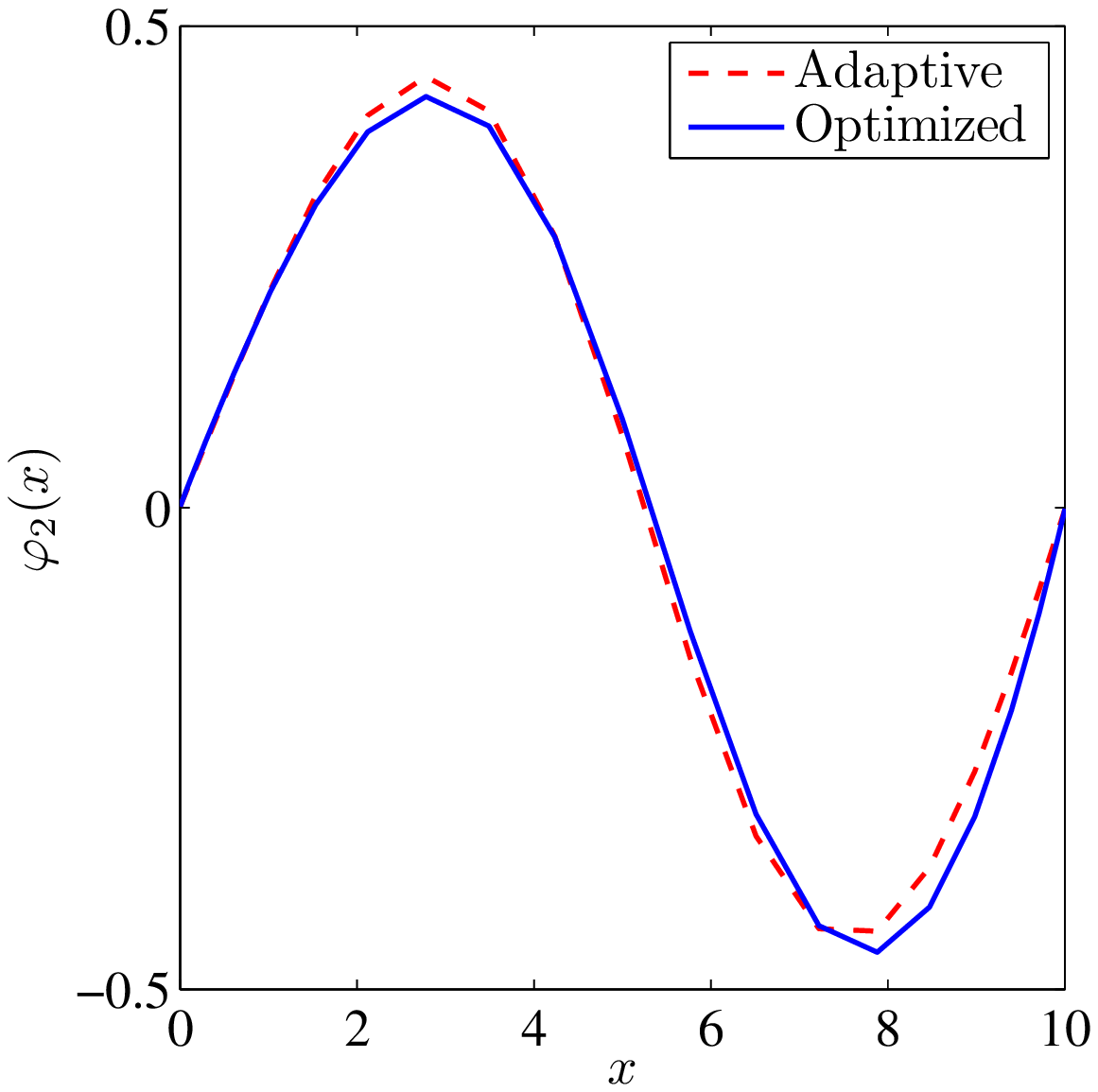}}

    \subfloat[$\phi_{3}$]{\includegraphics[width=0.35\textwidth,
    height=0.35\textwidth]{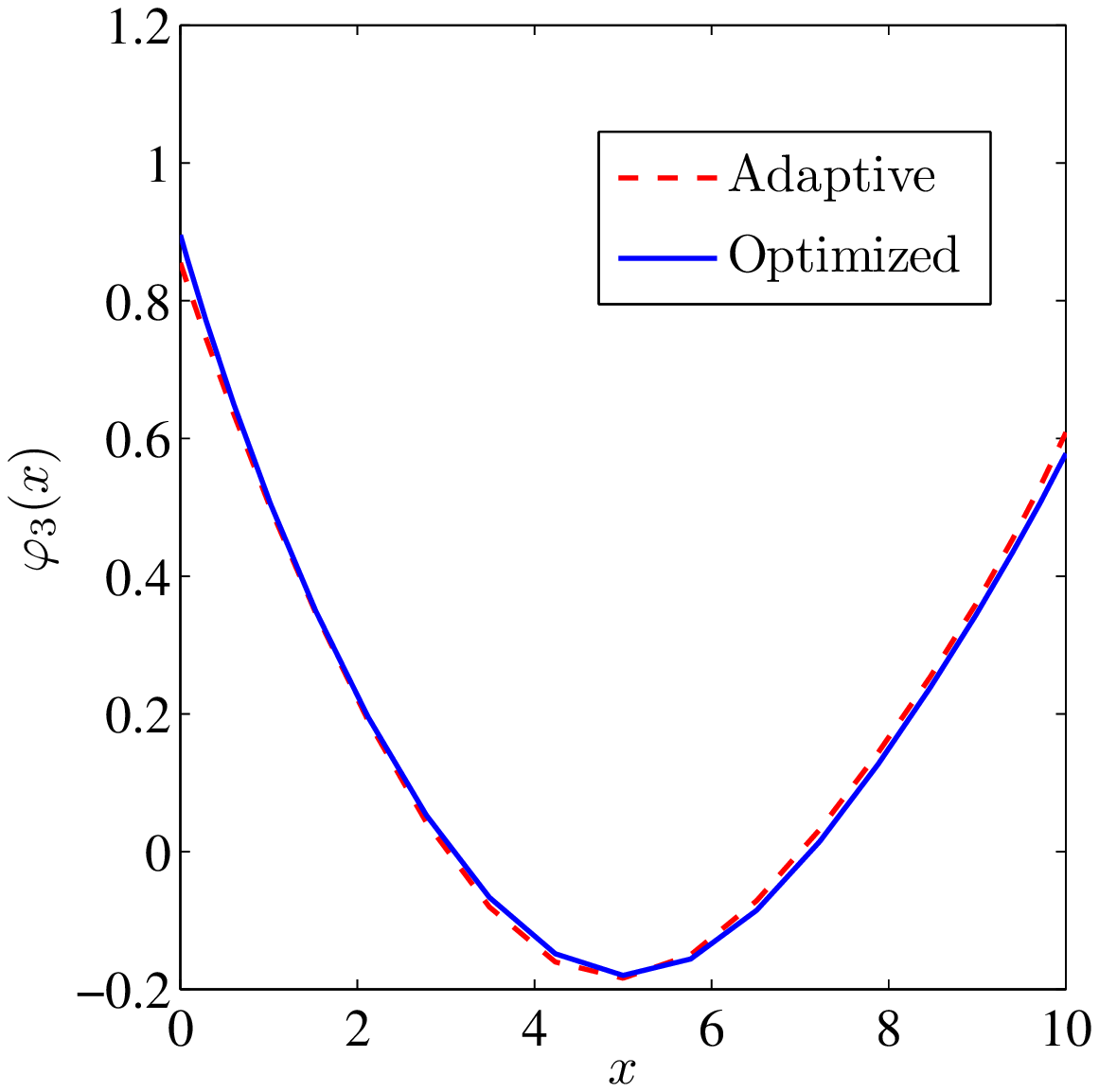}}
    \subfloat[$\phi_{4}$]{\includegraphics[width=0.35\textwidth,
    height=0.35\textwidth]{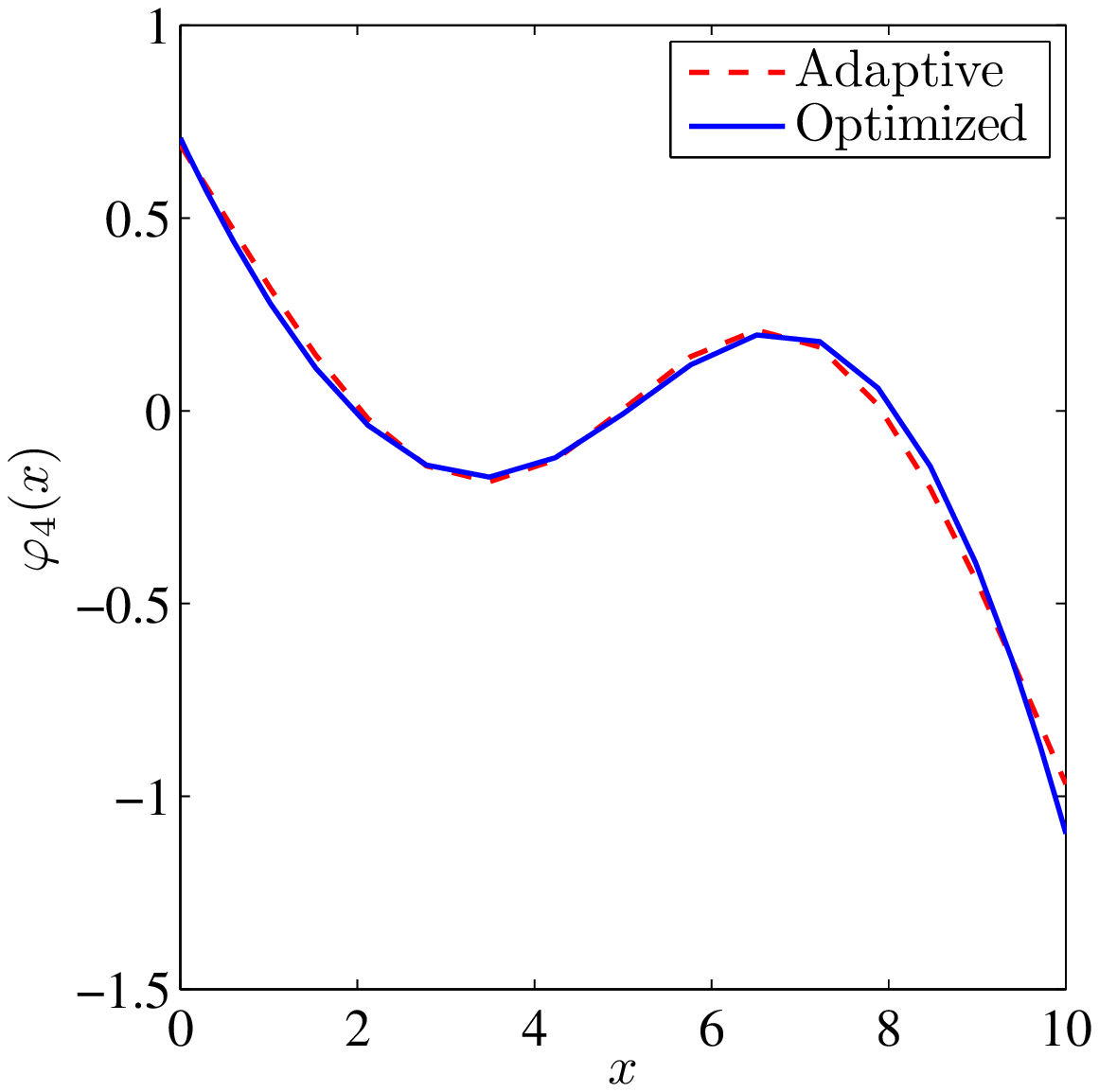}}
  \end{center}
  \caption{Comparison of the adaptive and optimized local basis
  functions for an insulating system  with $1$ electron per atom (spin
  neglected).  The adaptive local basis functions (red dashed line) and
  optimized local basis functions (blue solid line) are sorted according
  to the Ritz value of the local Hamiltonian in ascending order.}
  \label{fig:phiplot}
\end{figure}

For the insulating system, the relative error of the force is already
$0.5\%$ with as small as $4$ basis functions per atom using the
optimized local basis functions (Table~\ref{tab:ins1}).  When compared
to the adaptive local basis functions with the same number of basis
functions per atom, the error of the Helmholtz free energy per atom is
reduced by $51$ times, and the error of the force is reduced by $14$
times after the optimization procedure. It is illuminating to see the
difference between the adaptive local basis functions and the
optimized local basis functions.  Since any unitary transformation of
the basis functions in each element does not change the total energy
of the system, the basis functions should first be rotated according a
certain criterion.  Here we rotate the basis functions in an element
according to the Ritz values of the Hamiltonian in the same element.
Take the first element for example, the Hamiltonian operator is
denoted by $H_{11}$, and the basis functions in the first element is
denoted by $\Phi_1$. We solve the following eigenvalue problem
\begin{equation}
  (\Phi_{1}^T H_{11} \Phi_{1}) C_{1} = C_{1} \Lambda_{1},
  \label{}
\end{equation}
where $\Lambda_{1}$ is a diagonal matrix with values sorted in ascending
order.  Then we compare the rotated basis functions
\begin{equation}
  \Phi_{1}C \equiv [\varphi_{1},\cdots,\varphi_{J}]
  \label{}
\end{equation}
for adaptive and optimized local basis functions in
Fig.~\ref{fig:phiplot}. It is found that the optimized local basis
functions are very close to the adaptive local basis functions,
indicating that the adaptive local basis functions is already very
accurate in computing the total energy of the system.  The agreement
between the adaptive local basis functions and the optimized local
basis functions is very well for basis functions of low energy 
(Fig.~\ref{fig:phiplot} (a)), and the difference enlarges for basis
functions or higher energy. This can be understood as that the adaptive local basis
functions include contributions from unoccupied states with relatively
high energy level, while the optimized local basis functions reduce
the contribution from such unoccupied states by the optimization
procedure.

Similar results are found for metallic systems
(Table~\ref{tab:met1}).  More basis functions are needed in this case
since there are more electrons in the metallic system than those in
the insulating system studied here.  The relative error of the force
is $0.2\%$ with $8$ basis functions per atom using the optimized local
basis functions.  When compared to the adaptive local basis functions
using the same number of basis functions, the error of the Helmholtz
free energy per atom is reduced by $10$ times and the error of the
force is reduced by $60$ times using the optimized local basis
functions. The optimized local basis functions therefore greatly
improve the accuracy with the same number of basis functions.

\begin{table}[ht]
  \centering
  \begin{tabular}{c|c|c|c}
    \hline
    Method &  $\Delta \mc{F}_{\DG}$ / atom &  Absolute $\Delta F$   & Relative
    $\Delta F$ \\
    \hline
    Adaptive & $5.7\times 10^{-5}$ & $6.8\times 10^{-5}$ & $4.5\times 10^{-3}$\\
    \hline
    Optimized & $1.1\times 10^{-6}$ & $4.9\times 10^{-6}$ & $3.3\times 10^{-4}$\\
  \hline
  \end{tabular}
  \caption{Mean error of the Helmholtz free energy per atom, the absolute
    error of the force of the first atom, and the relative error of the
    force of the first atom.  This system is an insulating system  with
    $1$ electron per atom (spin neglected).  $4$ basis functions per atom
    are used for both the adaptive local basis functions and the
    optimized local basis functions.}
  \label{tab:ins1}
\end{table}

\begin{table}[ht]
  \centering
  \begin{tabular}{c|c|c|c}
    \hline
    Method &  $\Delta \mc{F}_{\DG}$ / atom &  Absolute $\Delta F$   & Relative
    $\Delta F$ \\
    \hline
    Adaptive & $1.4\times 10^{-3}$ & $3.8\times 10^{-4}$ & $9.6\times 10^{-2}$\\
    \hline
    Optimized & $1.7\times 10^{-4}$ & $4.5\times 10^{-6}$ & $1.6\times 10^{-3}$\\
  \hline
  \end{tabular}
  \caption{Mean error of the Helmholtz free energy per atom, the absolute
    error of the force of the first atom, and the relative error of the
    force of the first atom.  This system is a metallic system  with $4$ electrons
    per atom (spin neglected).  $8$ basis functions per atom are used for
    both the adaptive local basis functions and the optimized local
    basis functions.}
  \label{tab:met1}
\end{table}

On the other hand, the accuracy of using the adaptive local basis
functions can be systematically improved by increasing the number of
basis functions per atom.  For example, if the number of basis
functions per atom is increased from $8$ to $12$ for the metallic
system, the accuracy of using the adaptive local basis functions is
comparable to that of using the optimized local basis functions
(Table~\ref{tab:met2}).  This finding is fully consistent with the
previous work~\cite{LinLuYingE_Adaptive} that the adaptive local basis
functions also form an accurate and efficient local basis set for the
electronic structure calculation.  The mild increase of the number of
basis functions indicates that the adaptive local basis functions are
already very efficient at least for 1D or quasi-1D systems.  It is
also found in the previous work that the number of adaptive local
basis functions increases considerably from quasi-1D systems to 3D
bulk systems~\cite{LinLuYingE_Adaptive}.  We expect that the number of
basis functions can be reduced by a significant amount using optimized
local basis functions in 3D bulk systems.

\begin{table}[ht]
  \centering
  \begin{tabular}{c|c|c|c}
    \hline
    Method &  $\Delta \mc{F}_{\DG}$ / atom &  Absolute $\Delta F$   & Relative
    $\Delta F$ \\
    \hline
    Adaptive & $3.4\times 10^{-5}$ & $1.9\times 10^{-7}$ & $1.1\times 10^{-4}$\\
    \hline
    Optimized & $3.4\times 10^{-5}$ & $1.7\times 10^{-7}$ & $1.0\times 10^{-4}$\\
  \hline
  \end{tabular}
  \caption{Mean error of the Helmholtz free energy per atom, the absolute
    error of the force of the first atom, and the relative error of the
    force of the first atom.  This system is a metallic system with $4$ electrons
    per atom (spin neglected).  $12$ basis functions per atom are used for
    both the adaptive local basis functions and the optimized local
    basis functions.}
  \label{tab:met2}
\end{table}

We also test the optimized local basis functions on a system with
local defects. The defect system is obtained by
choosing the parameter $a$ at one atom in the
potential~\eqref{eqn:pot} to be different from the parameters $a$ of the
rest of the atoms. The system contains $8$ atoms with $4$
electrons and $8$ basis functions per atom.  The parameter $a$ is set
to be $5.0$ for all atoms except for the first atom which is
set to be $3.0$.  The error of the Helmholtz free energy per atom and
the error in the force of the defect atom are comparable to those in
the periodic case (Table~\ref{tab:defect1}).

\begin{table}[ht]
  \centering
  \begin{tabular}{c|c|c|c}
    \hline
    Method &  $\Delta \mc{F}_{\DG}$ / atom &  Absolute $\Delta F$   & Relative
    $\Delta F$ \\
    \hline
    Adaptive & $1.3\times 10^{-3}$ & $1.1\times 10^{-4}$ & $6.3\times 10^{-2}$\\
    \hline
    Optimized & $1.8\times 10^{-4}$ & $5.3\times 10^{-6}$ & $3.3\times 10^{-3}$\\
  \hline
  \end{tabular}
  \caption{Mean error of the Helmholtz free energy per atom, the absolute
    error of the force of the first atom, and the relative error of the
    force of the first atom for a metallic system with a defect.  $8$
    basis functions per atom are used for both the adaptive local basis
    functions and the optimized local basis functions.}
  \label{tab:defect1}
\end{table}

\begin{figure}[ht]
  \begin{center}
    \subfloat[Computational time]{\includegraphics[width=0.32\textwidth,
    height=0.32\textwidth]{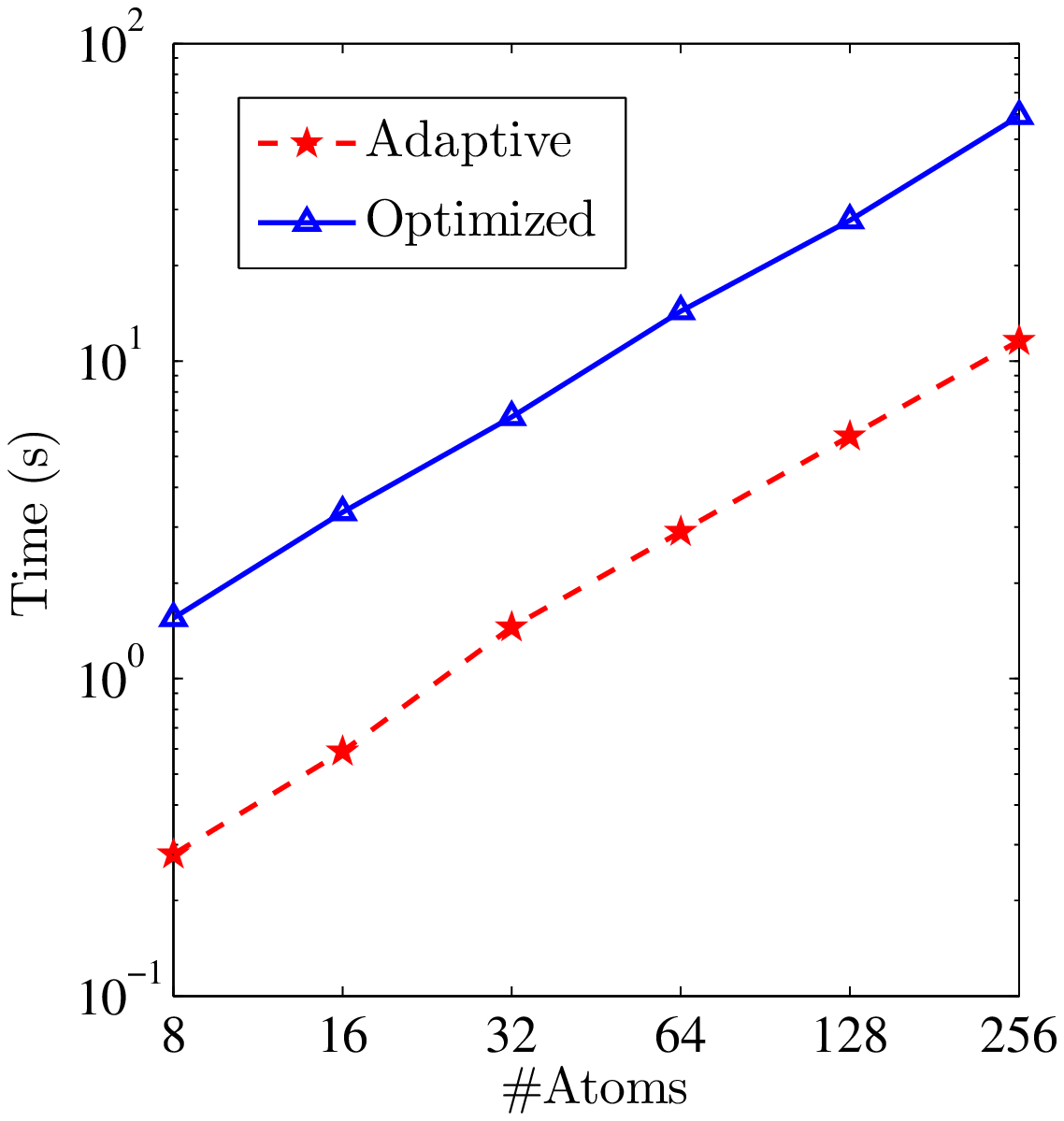}}
    \subfloat[Error of the Helmholtz free energy per atom]{\includegraphics[width=0.32\textwidth,
    height=0.32\textwidth]{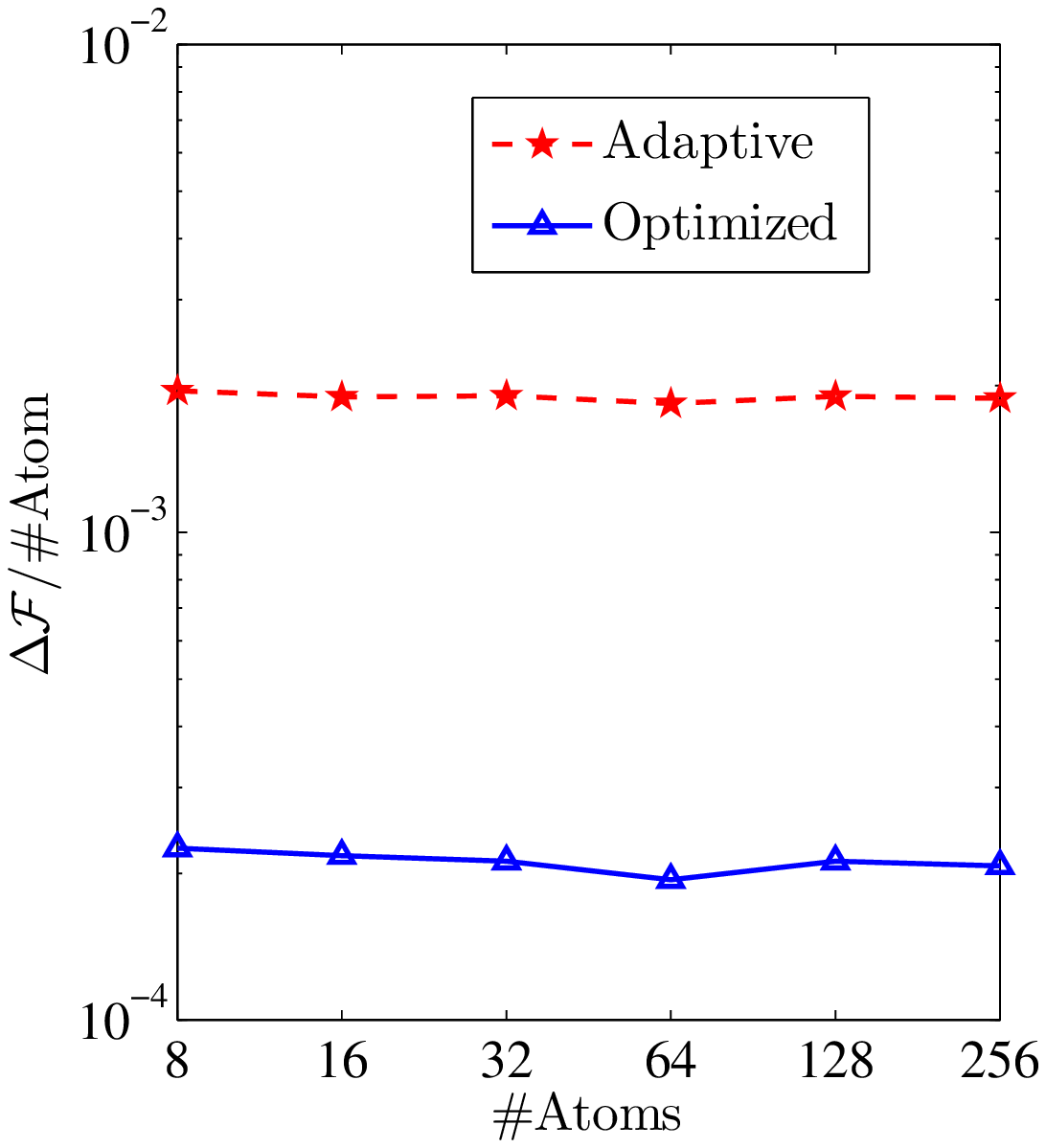}}
    \subfloat[Error of the force]{\includegraphics[width=0.32\textwidth,
    height=0.32\textwidth]{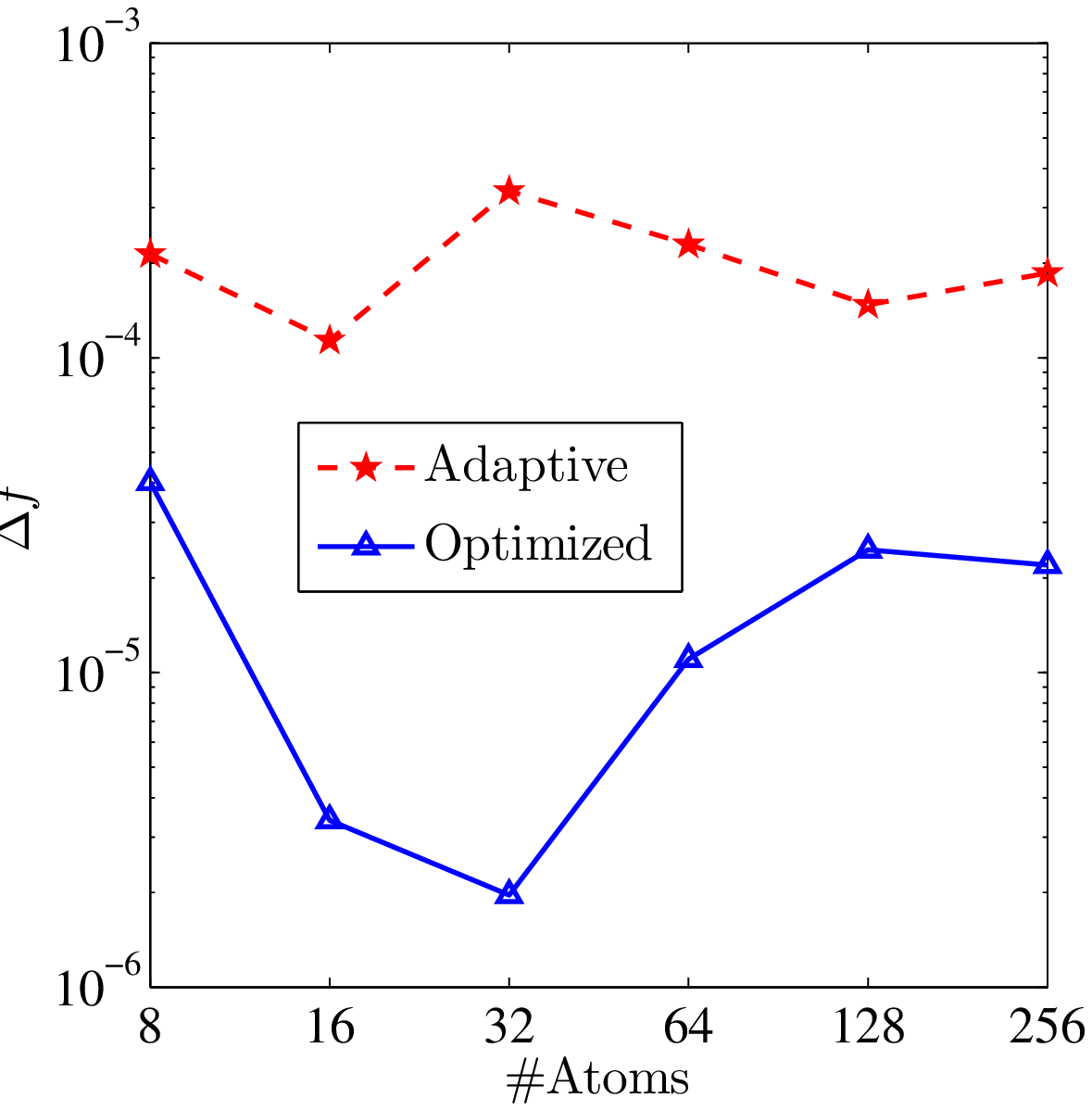}}
  \end{center}
  \caption{(a) The computational time for solving systems of various
  sizes using adaptive local basis functions (red dashed line) and 
  optimized local basis functions (blue solid line). 
  (b) The error of the Helmholtz free energy per atom using adaptive
  local basis functions (red dashed line) and optimized local basis
  functions (blue solid line).  (c) The absolute error of the force for
  the first atom using adaptive local basis functions (red dashed line)
  and optimized local basis functions (blue solid line).}
  \label{fig:large}
\end{figure}

Finally, we compare the performance of the adaptive local basis
functions and the optimized local basis functions for systems of
increasing size with $8,16,32,128,256$ atoms, respectively.  The
system is randomly perturbed by $0.2$ au from the crystalline
configuration, with a defect introduced at one atom of the
potential. The computational time for constructing the adaptive local
basis functions (red dashed line with star) and for constructing the
optimized local basis functions (blue solid line with triangle) are
compared in Fig.~\ref{fig:large} (a) plotted in logarithmic scale.
$5$ Newton steps and $30$ GMRES iterations are used for the outer
iteration and the inner iteration respectively in the optimization
procedure. Since the optimized local basis functions use the adaptive
local basis functions as an initial guess, the computational time for
the optimized local basis functions also includes that for the
adaptive local basis functions.  The computational time for
constructing both the adaptive local basis functions and the optimized
local basis functions are linear thanks to the locality of the basis
functions.  The construction of the optimized local basis functions is
$6\sim 9$ times more expensive than the construction of the adaptive
local basis functions, indicating that the optimization procedure
should be further improved in order to generate a practically
efficient optimized local basis set.  The error of the Helmholtz free
energy per atom and the error of the force on the first atom are shown
in Fig.~\ref{fig:large} (b) and (c), respectively. It is found that
the Helmholtz free energy obtained by the optimized local basis
functions is stably $8\sim 9$ times more accurate than that obtained
by the adaptive local basis functions.  The ratio of improvement of
the force has a much larger dependence on the realization of the
atomic configuration which ranges from $5\sim 170$ times, with the
average ratio of improvement being around one order of magnitude.

\subsection{Dynamic case}\label{subsec:dynamic}

The optimized local basis set is able to accurately compute the
electron energy and the force using a small number of basis functions.
Now we show that the optimized local basis functions can also be used
in molecular dynamics. We illustrate the performance of the optimized
local basis functions for molecular dynamics using the same metallic
system as in Section~\ref{subsec:static} with $4$ electrons and $8$
basis functions per atom.

In the Born-Oppenheimer approximation, the equations of motion for
atom $I$ are given by
\begin{equation}
  M_I \ddot{R}_I = F_I, \quad I=1,\cdots, N_A,
  \label{eqn:Newton}
\end{equation}
The mass of the ions $M_I$ is set to be $42000$ which is close the
mass of sodium in the atomic unit.  $F_{I}$ is the Hellman-Feynman
force in~\eqref{eqn:HF1D} for atom $I$. The equations of motion
\eqref{eqn:Newton} conserve the total energy given by
\begin{equation}
  E_{IC} = \sum_{I=1}^{N_A}\frac{M_I \dot{R}_I^2}{2} 
  + \mc{F}_{\DG}(R) + V_{II}(R).
  \label{eqn:conv}
\end{equation}
The numerical conservation of the total energy is quantified by the
drift of $E_{IC}$, which is defined as the relative difference of
$E_{IC}$ along the trajectory, \ie 
\begin{equation}
  \text{Drift}(t) = \frac{\abs{E_{IC}(t)-E_{IC}(0)}}{\abs{E_{IC}(0)}}.
  \label{eqn:drift}
\end{equation}

\begin{figure}[ht]
  \begin{center}
    \includegraphics[width=0.32\textwidth, height=0.32\textwidth]{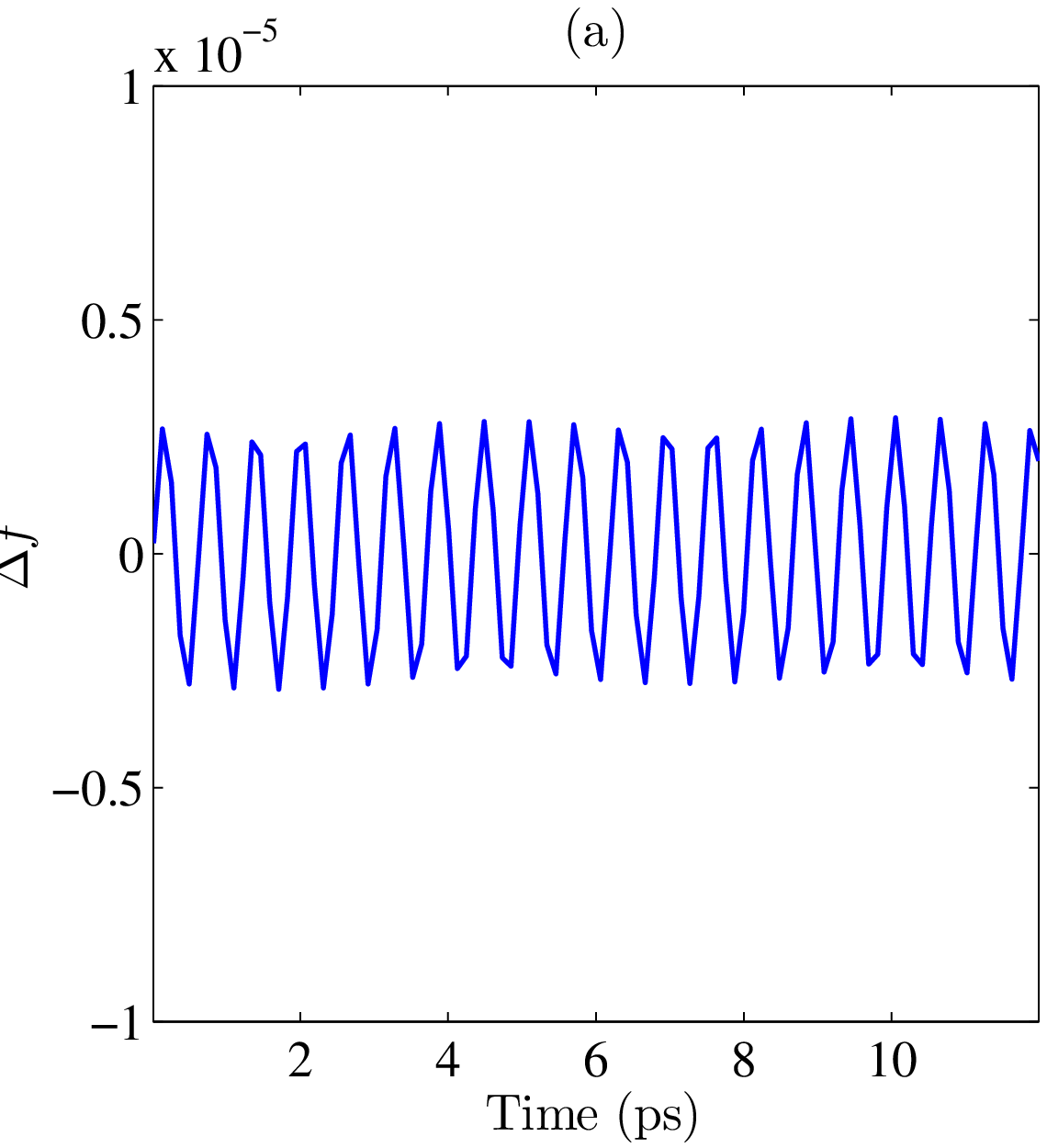}
    \includegraphics[width=0.32\textwidth, height=0.32\textwidth]{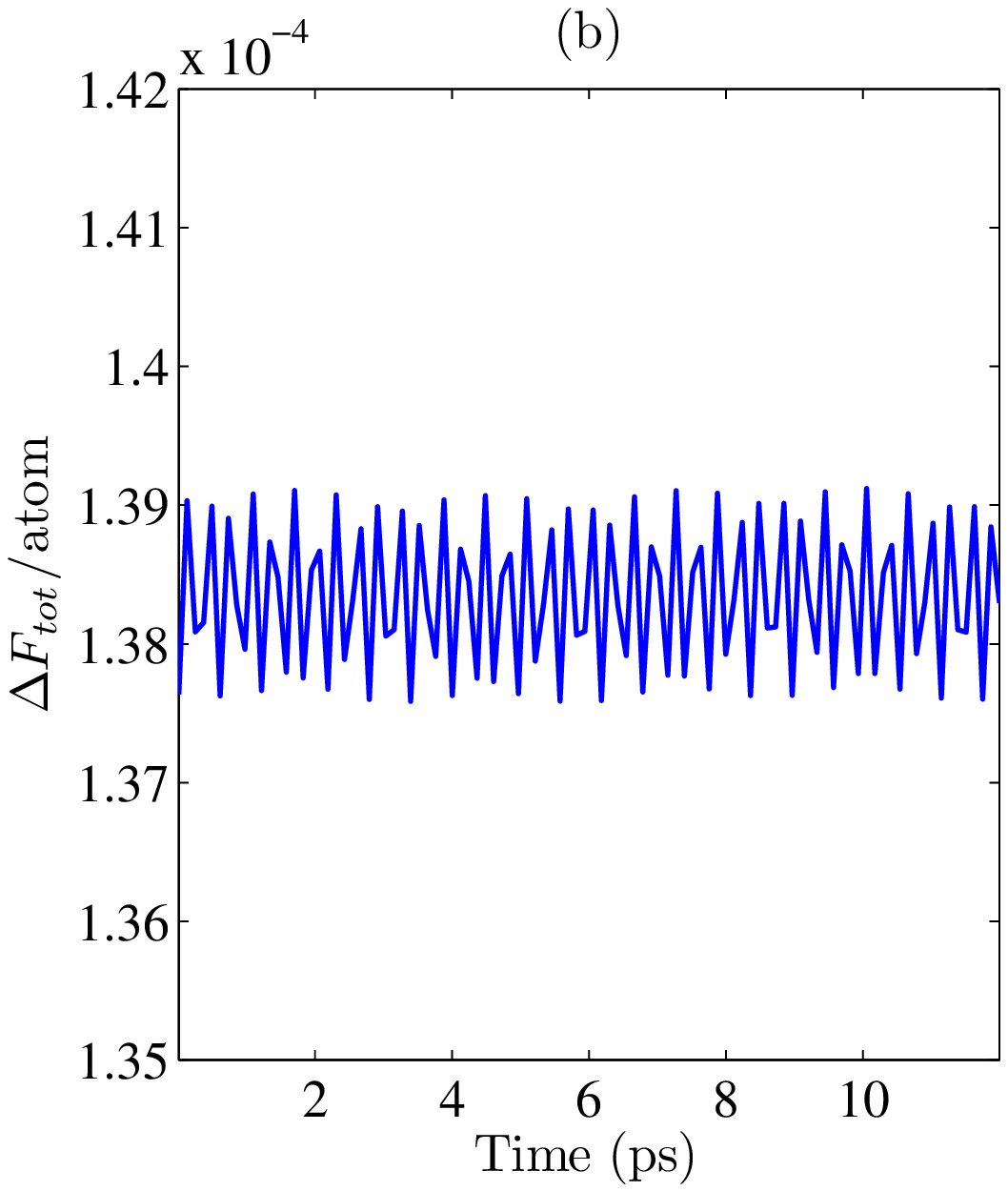}
    \includegraphics[width=0.32\textwidth, height=0.32\textwidth]{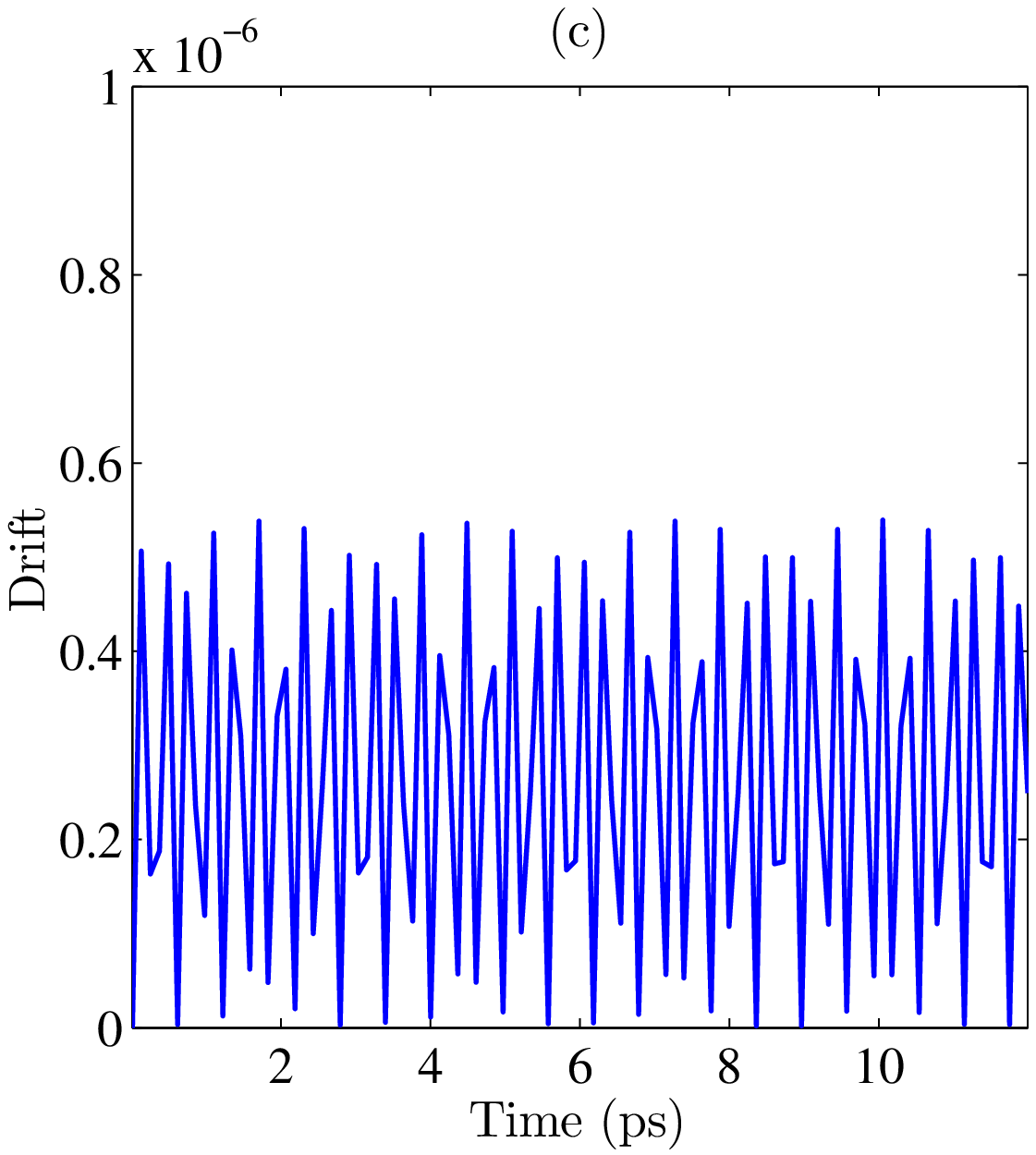}
  \end{center}
  \caption{The error of the force of the first atom (a), the error of
  the Helmholtz free energy per atom (b) and the drift of the conserved
  quantity (c) along the trajectory of the MD simulation plotted every
  $0.12$ ps.  The system is
  metallic with $4$ electrons and $8$ basis functions per atom.
  The mean deviation of the force is unbiased. }
  \label{fig:dynmet}
\end{figure}

Velocity-Verlet scheme~\cite{FrenkelSmit2002} is used to propagate the
equations of motion for the atoms with the time step $\Delta t=1.21$
femtoseconds (fs).  The simulation length is $10000$ steps and the
total length of the simulation is $12.1$ picoseconds (ps).  To ensure
the time-reversibility of the numerical scheme, the optimized local
basis functions use the adaptive local basis functions as the initial
guess at every time step.  However, this is not a necessary
requirement and can be improved by other time-reversible schemes such
as the extended Lagrangian Born-Oppenheimer
method~\cite{Niklasson2008}.  The initial configurations of the atoms
are perturbed by $0.2$ au away from the equilibrium equidistant
configuration, and the initial kinetic energy of the atoms is $1000$ K
with the mean velocity of all atoms (\ie the velocity of the centroid)
being zero.  The error of the force and the error of the Helmholtz
free energy per atom are well within $2.5 \times 10^{-6}$ and
$1.4\times 10^{-4}$, respectively (see Fig.~\ref{fig:dynmet} (a) and
(b)), which is consistent with the behavior of errors in the static
calculation.  The Helmholtz free energy obtained from the optimized
local basis functions is systematically higher than that in the
benchmark planewave simulation.  The sources of the systematic shift
are the penalty parameter $\alpha$ in the DG formulation, and that the
minimization procedure is restricted to an admissible set of the space
spanned by the primitive functions.  Nonetheless, the mean deviation
of the force is unbiased, indicating that the structure of the
trajectory obtained using the optimized local basis functions is well
preserved.  The drift of the conserved quantity~\eqref{eqn:drift} is
also well controlled within $5\times 10^{-7}$ (Fig.~\ref{fig:dynmet}
(c)).

% LL  Date: 08/02/2011
% Obtained by 
% C:\Users\linlin\Documents\researchBIN\optbasis\mdft\RUN8\plotfig.m

\section{Conclusion}\label{sec:conclusion}

We have developed the optimized local basis set to solve models in the
Kohn-Sham density functional theory for both insulating and metallic
systems. The optimized local basis functions form an accurate basis
set for computing the electron energy as well as the atomic force with
a small number of basis functions per atom.  When the optimality
condition is achieved, the optimized local basis functions give the
lowest energy among all the basis functions in an admissible set
determined by the primitive basis functions.  The force is accurately
described by the Hellmann-Feynman force, and the contribution of the
derivative of the basis functions (\ie the Pulay force) vanishes
automatically. The concept of the optimized local basis functions is
quite general, and the methods developed in this paper are useful for
other problems such as selecting basis functions and evaluating
parameter-dependent functions as well .

To obtain the optimized local basis functions in practice, we proposed
a preconditioned Newton-GMRES method. The resulting optimized local
basis functions are tested using a one-dimensional model problem.  We
find that the optimized local basis functions accurately compute the
Helmholtz free energy and the force using a very small number of basis
functions per atom for both insulating and metallic systems.  When
applied to the molecular dynamics simulation, the optimized local
basis functions do not exhibit any systematic drift in terms of the
force or the total energy for the ionic degrees of freedom.  Therefore
the optimized local basis functions are able to give the correct
statistical and dynamical properties along the molecular dynamics
trajectory, and can be used for long time molecular dynamics
simulation.

The optimized local basis set provides an implementable criterion
to eliminate the artificial effect in the force due to the change of the
basis functions and to maintain a small set of basis functions, which
makes the optimized local basis set an ideal tool in the molecular
dynamics simulation. However, the construction of the optimized local
basis functions is found to be already more expensive than other choices
such as adaptive local basis functions, indicating that the optimization
procedure should be further improved especially when applied to
Kohn-Sham density functional theory in 3D. The more efficient scheme may
be achieved by including a feasible approximation of the derivative of
the density matrix with respect to the basis function, a more efficient
preconditioner for the GMRES iteration, or even a more efficient gradient
method instead of a Newton-type method. These will be our future work.

\medskip
\noindent{\textbf{Acknowledgment:}}

W.~E and L.~L. are partially supported by DOE under Contract No.
DE-FG02-03ER25587 and by  NSF under Contract No. DMS-0914336.  L.~Y.
is partially supported by an Alfred P. Sloan Research Fellowship and an
NSF CAREER award DMS-0846501. The authors thank  the hospitality of
Shanghai Jiao Tong University where part of the work was done.

\appendix

\section{Finite temperature Kohn-Sham density functional theory}
\label{sec:dg}

In this appendix, we briefly described the basic formulation of the
Kohn-Sham density functional theory
\cite{HohenbergKohn:64,KohnSham:65} and its finite temperature
generalization. In the Kohn-Sham density functional theory, the
ground state electron energy is written as
\begin{multline}\label{eqn:KSfunc_0T}
  E_{\tot} = E_{\tot}(\{\psi_i\}) = \frac{1}{2} \sum_{i=1}^N \int \abs{\nabla
    \psi_i}^2 \ud x + \int V_{\ext} \rho \ud x
  + \sum_{\ell} \gamma_{\ell} \sum_{i=1}^N  \abs{\int b_{\ell}^{\ast} \psi_i \ud x}^2  \\
  + \frac{1}{2} \iint \frac{\rho(x) \rho(y)}{\abs{x-y}} \ud x \ud y +
  \int\exc[\rho(x)] \ud x,
\end{multline}
where the Kohn-Sham orbitals are the solutions to the minimization
problem
\begin{equation}\label{eqn:KSfunc_0T_min}
  \begin{split}
    &\min_{\{\psi_{i}\}_{i=1}^{N}} E_{\tot}(\{\psi_i\}),\\
    &\text{s.t.} \quad \int \psi_i^{\ast} \psi_j \ud x = \delta_{ij},\quad i,j=1,\cdots,N.
  \end{split}
\end{equation}
With slight abuse of the notation, we denote by $\{\psi_{i}\}$ both
the arguments in the minimization problem~\eqref{eqn:KSfunc_0T_min},
and the solutions to the minimization problem, \ie the Kohn-Sham
orbitals.  The electron density is $\rho(x) = \sum_{i=1}^{N}
\abs{\psi_i(x)}^2$.  We have neglected the spin degeneracy.  The first
term of~\eqref{eqn:KSfunc_0T} is the kinetic energy. The second
and third terms come from pseudo-potential, which we have taken the
Kleinman-Bylander form~\cite{KleinmanBylander:82}. The pseudopotential
is given by
\begin{equation*}
  V_{\mathrm{PS}} = V_{\ext} + \sum_{\ell} \gamma_{\ell}
  \ket{b_{\ell}}\bra{b_{\ell}}.
\end{equation*}
For each $\ell$, $b_{\ell}$ is a function supported locally in the
real space around the position of one of the atoms, $\gamma_{\ell} =
+1$ or $-1$, and we have used the Dirac bra-ket notation.  The fourth
term is the Coulomb interaction between electrons, and the fifth term
is the exchange-correlation functional, for which the local density
approximation (LDA)~\cite{CeperleyAlder1980,PerdewZunger1981} is
adopted. The proposed method can also be used for more complicated
exchange-correlation functionals such as the generalized gradient
approximation (GGA) functionals~\cite{PerdewBurkeErnzerhof1996_GGA}.

The ground state electron energy defined in~\eqref{eqn:KSfunc_0T} 
is applicable to insulating systems with large band gap, but is 
difficult to evaluate for zero-gap metallic systems.  For metallic
system, finite temperature KSDFT becomes the standard
tool~\cite{Mermin1965}, in which the Helmholtz free energy  is considered instead.
For given finite temperature $T>0$, the Helmholtz free
energy is given by
\begin{multline}\label{eq:Merminfunc}
  \mc{F}_{\tot} = \mc{F}_{\tot}(\{\psi_i\}, \{f_i\}) = \frac{1}{2} \sum_{i} f_i \int \abs{\nabla
    \psi_i}^2 \ud x + \int V_{\ext} \rho \ud x \\
    + \sum_{\ell}
  \gamma_{\ell} \sum_{i} f_i \abs{\int b_{\ell}^{\ast} \psi_i \ud x}^2 
  + \frac{1}{2} \iint \frac{\rho(x) \rho(y)}{\abs{x-y}} \ud x \ud y \\
  + \int\exc[\rho(x)] \ud x + \beta^{-1} \sum_i \bigl( f_i \ln f_i + (1
  - f_i) \ln(1 - f_i) \bigr).
\end{multline}
Correspondingly $\{\psi_{i}\}$ and $\{f_i\}$ are the solutions to the minimization
problem
\begin{equation}\label{eqn:KSfunc_finiteT}
  \begin{split}
    &\min_{\{\psi_{i}\},\{f_i\}} \mc{F}_{\tot}(\{\psi_i\},\{f_i\}),\\
    &\text{s.t.} \quad \int \psi_i^{\ast} \psi_j \ud x =
    \delta_{ij},\quad i,j=1,\cdots,\widetilde{N}.
  \end{split}
\end{equation}
Here $\beta$ is the inverse temperature $\beta = 1 / k_B T$.
The number of eigenstates $\widetilde{N}$ is chosen to be slightly larger
than the number of electrons $N$ in order to compensate for the finite
temperature effect, following the criterion that the occupation number
$f_{\widetilde{N}}$  is sufficiently  small (less than $10^{-8})$.
$\{f_i\} \in [0, 1]$ are the occupation numbers which add up to the
total number of electrons $N = \sum_{i=1}^{\widetilde{N}} f_i$, and the
electron density $\rho=\sum_{i=1}^{\widetilde{N}} f_i \abs{\psi_i}^2$.
Compared to~\eqref{eqn:KSfunc_0T}, the only extra term is the last
term, which characterizes the entropic contribution.

The Kohn-Sham equation, or the Euler-Lagrange equation associated with
\eqref{eqn:KSfunc_finiteT} reads 
\begin{equation}\label{eq:KSeqn}
  H[\rho] \psi_i = \Bigl( - \tfrac{1}{2} \Delta + V_{\eff}[\rho]
  + \sum_{\ell} \gamma_{\ell} \ket{b_{\ell}} \bra{b_{\ell}} \Bigr) \psi_i 
  = \lambda_i \psi_i,
\end{equation}
where the effective one-body potential $V_{\eff}$ is given by 
\begin{equation}\label{eq:Veff}
  V_{\eff}[\rho](x) = V_{\ext}(x) + \int \frac{\rho(y)}{\abs{x - y}}\ud y
  + \exc'[\rho(x)].
\end{equation}
The occupation numbers are given by
\begin{equation}\label{eq:FermiDirac}
  f_i = \frac{1}{1 + \exp( \beta(\lambda_i - \mu))},
\end{equation}
which is the Fermi-Dirac distribution evaluated at $\lambda_{i}$. Here $\mu$ is the chemical
potential, which is chosen so that $f_i$ satisfies
\begin{equation}\label{eq:ficonstraint}
  \sum_i f_i = N.
\end{equation}
Note that \eqref{eq:KSeqn} is a nonlinear eigenvalue problem, as
$V_{\eff}$ depends on $\rho$, which is in turn determined by
$\{\psi_i\}$.  The electron density is self-consistent if
both~\eqref{eq:KSeqn} and~\eqref{eq:Veff} are satisfied. After
obtaining the self-consistent electron density, the Helmholtz free
energy can be expressed as
\begin{multline}\label{eq:Ftot1} 
  \mc{F}_{\tot}=\mc{F}_{\tot}(\rho, \mu) = \sum_{i} f_i \lambda_{i} + \beta^{-1} \sum_i \bigl(
  f_i \ln f_i + (1 - f_i) \ln(1 - f_i) \bigr) \\
  - \frac12 \iint \frac{\rho(x) \rho(y)}{\abs{x-y}} \ud x \ud y +
  \int\exc[\rho(x)] \ud x - \int \exc'[\rho(x)] \rho(x) \ud x.
\end{multline}
The goal of finite temperature Kohn-Sham density functional theory is
to calculate the free energy $\mc{F}_{\tot}$, the self-consistent
electron density $\rho$ and also the chemical potential $\mu$ given
the number of electrons, the temperature and the atomic configuration.
The Helmholtz free energy $\mc{F}_{\tot}(R)$ plays the role of the
electron energy $E(R)$ in Section~\ref{sec:intro}, and the force is
defined as the negative gradient of the Helmholtz free energy
$F(R)=-\frac{\partial \mc{F}_{\tot}(R)}{\partial R}$.  The Helmholtz
free energy is applicable to both the insulating and the metallic
systems.  As $T\to 0$, the Helmholtz free energy $\mc{F}_{\tot}$
reduces to the ground state electron energy $E_{\tot}$.
Therefore~\eqref{eqn:KSfunc_0T} is also called the zero temperature
KSDFT.

As $f_i$ is given by the Fermi-Dirac distribution, we have
\begin{align}
  & \sum_{i} f_i \lambda_{i} = \Tr \frac{H}{1 + \exp( \beta (H - \mu))}; \\
  & \sum_i f_i \ln f_i = \Tr \frac{1}{1+\exp(\beta(H - \mu))}
  \ln  \frac{1}{1+\exp(\beta(H - \mu))}; \\
  & \sum_i (1 - f_i) \ln(1 - f_i) = \Tr \frac{\exp(\beta(H -
    \mu))}{1+\exp(\beta(H - \mu))} \ln \frac{\exp(\beta(H -
    \mu))}{1+\exp(\beta(H - \mu))}.
\end{align}
Using these, we can rewrite \eqref{eq:Ftot1}
as (see \textit{e.g.} \cite{AlaviKohanoffParrinelloEtAl1994})
\begin{multline}\label{eq:Ftot2}
  \mc{F}_{\tot}(\rho, \mu) = - \beta^{-1} \Tr \ln ( 1+ \exp(\beta( \mu -
  H[\rho]))) + \mu N  \\
  - \frac12 \iint \frac{\rho(x) \rho(y)}{\abs{x-y}} \ud x \ud y +
  \int\exc[\rho(x)] \ud x - \int \exc'[\rho(x)] \rho(x) \ud x.
\end{multline}
One can verify by straightforward calculations that
\begin{equation}
  \frac{\delta \mc{F}_{\tot}(\rho, \mu)}{\delta \rho} = 0
\end{equation}
if $\rho$ and $\mu$ are the self-consistent solution of the
Kohn-Sham equation~\eqref{eq:KSeqn}. 
Taking derivative of \eqref{eq:Ftot2} with respect to $\mu$, we have
\begin{equation}
  \frac{\partial \mc{F}_{\tot}(\rho, \mu)}{\partial \mu}
  = - \Tr \frac{\exp(\beta(\mu - H))}{1 + \exp(\beta(\mu - H))} + 
  N = 0.
\end{equation}
Therefore, the atomic force takes the form
\begin{equation}
  \begin{aligned}
    F & = - \frac{\ud \mc{F}_{\tot}(\rho, \mu, R)}{\ud R} = -
    \frac{\partial \mc{F}_{\tot}(\rho, \mu, R)}{\partial R}  \\
    & = - \Tr \left[\frac{1}{1 + \exp(\beta(H - \mu))} \frac{\partial
      H}{\partial R}\right].
  \end{aligned}
\end{equation}
This is known as the Hellman-Feynman theorem at finite temperature.

The Kohn-Sham density functional theory is usually solved by using the
self-consistent iteration, where at each iteration, the electron
density $\wt{\rho}$ is obtained from effective Hamiltonian $H_{\eff}$.
Given an effective potential $V_{\eff}$, and hence the effective
Hamiltonian
\begin{equation}  
  H_{\eff} = - \tfrac{1}{2} \Delta + V_{\eff} + 
  \sum_{\ell} \gamma_{\ell} \ket{b_{\ell}} \bra{b_{\ell}},
\end{equation}
we find $\wt{\rho}$ from $\wt{\rho}(x) = \sum_i f_i \abs{\psi_i(x)}^2$
where $\{\psi_i\}$'s are eigenfunctions of $H_{\eff}$, and the
definition of $\{f_i\}$ follows~\eqref{eq:FermiDirac} and
\eqref{eq:ficonstraint}. Note that the $\{\psi_i\}$ and $\{f_i\}$'s
minimize the variational problem
\begin{multline}\label{eq:linearvar}
  \mc{F}_{\eff}(\{\psi_i\}, \{f_i\}) = \frac{1}{2} \sum_{i} \int f_i
  \abs{\nabla \psi_i(x)}^2 \ud x + \int V_{\eff}(x) \rho(x) \ud x \\
  + \sum_{\ell} \gamma_{\ell} \sum_{i} f_i \abs{ \langle b_{\ell},
    \psi_i \rangle }^2 + \beta^{-1} \sum_i \bigl( f_i \ln f_i +
  (1-f_i) \ln (1 - f_i) \bigr),
\end{multline}
with the orthonormality constraints $\langle \psi_i \vert \psi_j \rangle =
\delta_{ij}$.  

% We note that although Eq.~\eqref{eq:Merminfunc} and
% Eq.~\eqref{eq:linearvar} shares similarity in their expressions,
% Eq.~\eqref{eq:Merminfunc} is a general nonlinear optimization problem
% while Eq.~\eqref{eq:linearvar} is a quadratic optimization problem.
% Therefore Eq.~\eqref{eq:linearvar} is a much easier to solve.

%\bibliographystyle{elsarticle-num} 
%\bibliography{optdft}

\end{document}